\documentclass[twocolumn]{aastex701}
\usepackage{amsmath}
\usepackage{CJK}
\usepackage{makecell}
\usepackage{multirow}

\newcommand{\omc}{$\omega$\,Cen}

\newcommand{\vlos}{$V_{\rm LOS}$}

\newcommand{\perpixel}{$\text{pix}^{-1}$}

\newcommand{\snrmuse}{$\rm{S/N}_{\rm{MUSE}}$}
\newcommand{\teff}{\ensuremath{T_{\mathrm{eff}}}}
\newcommand{\feh}{\ensuremath{[\mathrm{Fe/H}]}}
\newcommand{\logg}{\mbox{$\log g$}}
\newcommand{\xfe}{\ensuremath{[\mathrm{X/Fe}]}}

\newcommand{\nfe}{\ensuremath{[\mathrm{N/Fe}]}}

\newcommand{\mgfe}{\ensuremath{[\mathrm{Mg/Fe}]}}

\newcommand{\nad}{\ensuremath{\mathrm{Na\,\textsc{i}\,D}}}

\newcommand{\ddpayne}{\textsc{DD-Payne}}
\newcommand{\ddpayneg}{\textsc{DD-Payne-G}}
\newcommand{\ddpaynea}{\textsc{DD-Payne-A}}

\newcommand{\thepayne}{\textsc{The Payne}}

\newcommand{\abinitio}{\emph{ab initio}}

\defcitealias{Nitschai2023ApJ}{N23}
\defcitealias{Haberle2024ApJ}{H24}
\defcitealias{Xiang2019ApJS}{X19}

\received{XX}
\revised{XX}
\accepted{XX}

\submitjournal{APJ}

\begin{document}
\begin{CJK*}{UTF8}{gbsn}
\title{oMEGACat. IX. Chemical Tagging of Omega Centauri Populations with Machine-Learning-Inferred Abundances from the MUSE Spectrograph}

\author[0000-0003-2512-6892]{Z. Wang (王梓先)}
\affiliation{Department of Physics and Astronomy, University of Utah, Salt Lake City, UT 84112, USA}
\email[show]{wang.zixian.astro@gmail.com}
\correspondingauthor{Zixian Wang (Purmortal, 王梓先)}

\author[0000-0003-0248-5470]{A. C. Seth}
\affiliation{Department of Physics and Astronomy, University of Utah, Salt Lake City, UT 84112, USA}
\email{aseth@astro.utah.edu}

\author[0009-0005-8057-0031]{C. Clontz}
\affiliation{Department of Physics and Astronomy, University of Utah, Salt Lake City, UT 84112, USA}
\affiliation{Max Planck Institute for Astronomy, K\"onigstuhl 17, D-69117 Heidelberg, Germany}
\email{clontz@mpia.de}

\author[0000-0002-6922-2598]{N. Neumayer}
\affiliation{Max Planck Institute for Astronomy, K\"onigstuhl 17, D-69117 Heidelberg, Germany}
\email{neumayer@mpia.de}

\author[0000-0002-5844-4443]{M. H\"aberle}
\affiliation{European Southern Observatory, Karl-Schwarzschild-Straße 2, 85748 Garching, Germany}
\email{maximilian.haberle@eso.org}

\author[0000-0001-6604-0505]{S. Kamann}
\affiliation{Astrophysics Research Institute, Liverpool John Moores University, 146 Brownlow Hill, Liverpool L3 5RF, UK}
\email{S.Kamann@ljmu.ac.uk}

\author[0000-0002-7547-6180]{M. Latour}
\affiliation{Institut f\"ur Astrophysik und Geophysik, Georg-August-Universit\"at G\"ottingen, Friedrich-Hund-Platz 1, 37077 G\"ottingen, Germany}
\email{marilyn.latour@uni-goettingen.de}

\author[0000-0002-2941-4480]{M. S. Nitschai}
\affiliation{Max Planck Institute for Astronomy, K\"onigstuhl 17, D-69117 Heidelberg, Germany}
\email{nitschai@mpia.de}

\author[0000-0002-7489-5244]{P. J. Smith}
\affiliation{Max Planck Institute for Astronomy, K\"onigstuhl 17, D-69117 Heidelberg, Germany}
\affiliation{Fakult\"at f\"ur Physik und Astronomie, Universit\"at Heidelberg, Im Neuenheimer Feld 226, D-69120 Heidelberg, Germany}
\email{pesmith@mpia.de}

\author[0000-0001-8052-969X]{S. O. Souza}
\affiliation{Max Planck Institute for Astronomy, K\"onigstuhl 17, D-69117 Heidelberg, Germany}
\email{s-souza@mpia.de}

\author[0000-0003-4546-7731]{G. van de Ven}
\affiliation{Department of Astrophysics, University of Vienna, T\"urkenschanzstrasse 17, 1180 Wien, Austria}
\email{glenn.vandeven@univie.ac.at}

\author[0000-0002-1212-2844]{M. Alfaro-Cuello}
\affiliation{Centro de Investigaci\'{o}n en Ciencias del Espacio y F\'{i}sica Te\'{o}rica, Universidad Central de Chile, La Serena 1710164, Chile}
\email{mayte.alfaro@ucentral.cl}

\author[0000-0003-3858-637X]{A. Bellini}
\affiliation{Space Telescope Science Institute, 3700 San Martin Dr., Baltimore, MD, 21218, USA}
\email{bellini@stsci.edu}

\author[0000-0002-0160-7221]{A. Feldmeier-Krause}
\affiliation{Department of Astrophysics, University of Vienna, T\"urkenschanzstrasse 17, 1180 Wien, Austria}
\email{anja.krause@univie.ac.at}

\author[0000-0002-6072-6669]{N. Kacharov}
\affiliation{Leibniz Institute for Astrophysics, An der Sternwarte 16, 14482 Potsdam, Germany}
\email{kacharov@aip.de}

\author[0000-0001-9673-7397]{M. Libralato}
\affiliation{INAF - Osservatorio Astronomico di Padova, Vicolo dell'Osservatorio 5, Padova I-35122, Italy}
\email{mattia.libralato@inaf.it}

\author[0000-0001-7506-930X]{A. P. Milone}
\affiliation{Dipartimento di Fisica e Astronomia “Galileo Galilei”, Universita’ di Padova, Vicolo dell’Osservatorio 3, Padova, IT-35122}
\affiliation{Istituto Nazionale di Astrofisica - Osservatorio Astronomico di Padova, Vicolo dell’Osservatorio 5, Padova, IT-35122}
\email{antonino.milone@unipd.it}

% \author{the oMEGACat Collaboration}
% \noaffiliation



\begin{abstract}
We present chemical abundance measurements for 7,302 red giant branch stars within the half-light radius ($\sim5'$) of $\omega$~Centauri (\omc{}), derived from MUSE spectra using the neural network model \ddpayne{}.
\ddpayne{} effectively identifies spectral features of C, N, and O for $\feh{}>-1.0$~dex; Mg for $\feh{}>-1.5$~dex; and Na, Ca, and Ba for all metallicities. %($\feh{}<-0.5$~dex).
By combining these measurements with previous high-resolution studies, we create the most comprehensive picture of \omc{}'s rich chemical evolutionary history.
For the first time, we map elemental variations across the entire chromosome diagram, which is widely used to identify multiple populations.
We analyze the median chemical abundance trends as functions of age and metallicity for different subpopulations.
The \ddpayne{} measurements of [C/Fe], [N/Fe], and [O/Fe] extend literature trends to higher metallicities and show continuous abundance-metallicity relations, with [(C+N+O)/Fe] increasing steadily with \feh{}.
[Ca/Fe] and the $s$-process element [Ba/Fe] also increase with metallicity across all populations. For [Ba/Fe], the chemically enhanced (P2) populations are more enriched than primordial (P1) and the intermediate (Im) populations.
Furthermore, [N/Fe] correlates strongly with stellar age while [Ca/Fe] and [Ba/Fe] exhibit a weaker age dependence.
% The [Ba/Ca] ratio suggests stronger neutron-capture enrichment than core-collapse supernovae production.
% Using these abundance-metallicity-age relations, we evaluate different formation scenarios of \omc{} proposed in the literature.
Using these abundance-metallicity-age relations, we evaluate formation scenarios of \omc{} proposed in the literature and find none fully explains all the observed relations.
Our study demonstrates that combining MUSE with machine learning enables large-sample stellar abundance measurements in crowded cluster cores, overcoming the limitations of fiber-fed spectroscopy for studying multiple stellar populations and their evolutionary histories.
\end{abstract}



\section{Introduction} 
\label{sec:intro}

% Brief introduction of omega cen
% Globular clusters provide unique windows into the chemical enrichment of compact stellar systems in the early Universe.
Omega Centauri (\omc{}; NGC~5139), the most massive stripped nuclear star cluster (NSC) in the Milky Way, remains one of the most challenging stellar systems to unravel in terms of its formation and evolution.
Unlike typical mono-metallic globular clusters, \omc{} exhibits a large metallicity spread of $>1$~dex \citep[e.g.,][]{Freeman1975ApJL, Norris1995ApJ, Johnson2010ApJ, Nitschai2024ApJ} and hosts multiple stellar populations \citep[e.g.,][]{Bedin2004ApJL, Bellini2017ApJ, Latour2021A&A, Jofre2025A&A, Clontz2026ApJ} with distinct chemical abundance patterns \citep[e.g.,][]{Norris1995ApJ, Origlia2003ApJ, Marino2011ApJ, Gratton2011A&A}.
An intermediate-mass black hole (IMBH) has also been recently confirmed in \omc{} through the detection of fast-moving stars \citep{Haberle2024Natur}, further supporting its origin as the stripped nucleus of an accreted dwarf galaxy \citep[e.g.,][]{Bekki2003MNRAS, Tsuchiya2003ApJL, Ideta2004ApJL, Majewski2012ApJL}.
All these features point to a highly complex formation history of \omc{}, which therefore makes it a key system both for understanding the formation of nuclear star clusters \citep{Neumayer2020A&ARv} and the assembly history of the Milky Way \citep[e.g.,][]{Xiang2022Natur, Anguiano2025ApJ, Pagnini2026A&A}.

% \textcolor{red}{Mention the population identification by metallicity: Pancino et al. (2000) who used wide-field photometry to clearly resolve the Red Giant Branch (RGB) into distinct subpopulations, specifically defining:
% RGB-MP (Metal-Poor, the dominant population).
% RGB-MInt (Metal-Intermediate).
% RGB-a (Anomalous, the most metal-rich tail).
% This classification established the "Intermediate" group as a structurally distinct population rather than just a continuous spread. Later spectroscopic studies, such as Johnson \& Pilachowski (2010), further refined this into subgroups (RGB-Int1, RGB-Int2, RGB-Int3) based on detailed chemical abundances.}

% Photometric evidence for multiple populations
Decades of photometric studies have characterized \omc{}'s multiple populations. 
Early observations found multiple sequences in the cluster's color-magnitude diagram \citep[e.g.,][]{Norris1975ApJL, Lee1999Natur, Pancino2000ApJL, Ferraro2004ApJL}. 
Notably, \omc{} hosts a split main-sequence, with a blue branch that is more helium-rich (and slightly more metal-rich) than the red branch \citep{Bedin2004ApJL, Piotto2005ApJ, King2012AJ, Latour2021A&A}.  
In recent years, Hubble Space Telescope (\textit{HST}) multi-band photometry has enabled the construction of ``chromosome diagrams''; pseudo-color-color diagrams that disentangle multiple stellar populations within star clusters based on variations in light elements (He, N, O) \citep[e.g., ][]{Sbordone2011A&A, Milone2017MNRASa, Milone2017MNRASb, Marino2019MNRAS}.  
The chromosome map of \omc{} reveals at least three primary populations separated vertically in the diagram. On the bottom are stars with field-like abundances, a.k.a., P1 or first-generation stars. Enriched stars fall on top two separate sequences, an intermediate and upper sequence (Im and P2 hereafter) that are also often referred to as second-generation stars 
%the first- and second-generation stars (P1 and P2) that lie on the upper and lower $\Delta_{C}$ values, and an ``intermediate'' (Im) population in between 
\citep{Nitschai2024ApJ, Clontz2025ApJ, Mason2026MNRAS, Dondoglio2026A&A}.
A more detailed analysis of the chromosome diagram with the application of clustering algorithms also revealed that these primary groups of stars can be further divided into $\sim15$ subpopulations \citep{Bellini2017ApJ, Clontz2026ApJ}.
These photometric subpopulations indicate the presence of multiple episodes of star formation and chemical enrichment within \omc{}.

% Age
% \omc{} also has an extended star formation history compared to other type I globular clusters.
Early analysis of sub-giant branch (SGB) stars suggested a prolonged star formation history, with the inferred age spread ranging from 2 to 4~Gyr \citep[e.g.,][]{Stanford2006ApJ,Joo2013ApJ, Villanova2014ApJ, Tailo2016MNRAS}. 
\citet{Clontz2024ApJ} combined metallicity and photometry for 8100 SGB stars \citep{Nitschai2023ApJ, Haberle2024ApJ}, with new isochrones that include the impact of helium and C+N+O to construct a high-precision age-metallicity relation with most stars ranging in age from 10 to 13 Gyr.  
%\citet{Joo2013ApJ} included specific effects of helium and CNO enhancement and narrowed the age dispersion to $\sim1.5$~Gyr. 
%Most recently, \citet{Clontz2024ApJ, Clontz2025ApJ} utilized the largest spectroscopic sample of SGB stars to date from the oMEGACat dataset \citep{Nitschai2023ApJ, Haberle2024ApJ} and constructed a high-precision age-metallicity relation. 
They resolved two distinct formation sequences, with a tighter, lower metallicity sequence and a more diffuse, higher metallicity population.  
By combining these age measurements with subpopulation clustering, \citet{Clontz2026ApJ} found that P2 stars are $\sim$1~Gyr younger than P1 stars, with the Im population falling in between.

% Spectroscopic abundance measurements
Spectroscopic studies have provided direct chemical abundance measurements and confirmed their complex variations in \omc{}. 
% helium
One of the most striking deviations of \omc{} is its extreme helium enrichment.
A population with $Y\simeq0.39-0.44$ has been identified in contrast to the primordial helium population ($Y\simeq0.22$) \citep{Dupree2011ApJ, Dupree2013ApJL, Hema2014ApJL, Hema2018ApJ, Hema2020ApJ, Clontz2025ApJ}.
\citet{Clontz2025ApJ} confirmed that these helium-enriched stars correspond directly to the P2 population on the chromosome diagram.
Notably, significant helium enrichment is observed even at the lowest metallicities.
% light: CNO, Mg, Al, Na
In terms of light elements, \omc{} exhibits the classical anti-correlations observed in many globular clusters, but with unique characteristics.
High-resolution spectroscopic surveys have mapped a Na-O anti-correlation across all metallicities \citep{Johnson2010ApJ, Marino2011ApJ}, along with an Mg-Al anti-correlation that is particularly extended in the metal-poor and intermediate populations \citep{Meszaros2021MNRAS, AlvarezGaray2024AA}. 
Furthermore, measurements of C, N, and O have revealed a pattern that is distinct from the constant C+N+O sum typically found in mono-metallic clusters \citep[e.g.,][]{Meszaros2020MNRAS}. 
\cite{Marino2011ApJ, Marino2012ApJ} found that the total [(C+N+O)/Fe] abundance increases by approximately 0.5 dex from the most metal-poor component to the metal-rich population, with similar values at the same metallicity in both P1 and P2 populations. 
%This implies that the gas fueling subsequent generations was processed by advanced CNO cycling.
% Heavy: r/s-process
%For heavy elements, the most metal-poor stars are dominated by rapid neutron-capture ($r$-process) enrichment \citep{Norris1995ApJ, Smith2000AJ, Johnson2010ApJ}. 
Similarly, slow neutron capture ($s$-process) elements such as Ba and La increase monotonically with metallicity to extremely high values \citep[e.g.][]{Johnson2010ApJ, Marino2011ApJ, Marino2012ApJ}.
% , with no significant difference observed between P1 and P2 at the same metallicity \citep{Marino2011ApJ}. 

% Contradictory studies in interpreting omega cen formation history
Despite the above detailed studies on \omc{}'s multiple populations with age spread and abundance variations, the formation and evolution history of \omc{} remain elusive and heavily debated.
For example, \citet{Dondoglio2026A&A} argued for a self-enrichment scenario based on a combination of enrichment patterns and the radial profiles of the P2 stars being consistently more concentrated than the P1 population. % (although more centrally concentrated), 
They suggest that P2 formed directly from the ejecta of P1 stars, and that the intermediate (Im) population formed later from the combined ejecta of both P1 and P2. 
In contrast, \citet{Mason2026MNRAS} proposed a hierarchical assembly scenario based on chemical tagging with observations from APOGEE DR17 \citep{Abdurro'uf2022ApJS}. They suggested that the Im population may instead originate from gas-rich globular clusters that spiraled into the main \omc{} progenitor, triggering the starburst that produced the P2 population. 
Resolving these contradictory interpretations requires a direct linking of photometrically-identified populations with detailed ages and large samples of chemical abundance measurements.
However, previous spectroscopic studies of \omc{} have mostly relied on fiber-fed instruments (e.g., FLAMES, APOGEE), which suffer from contamination of the fiber light by adjacent stars in dense regions.
As a result, detailed abundance analyses have been restricted to $<2000$ stars in the outskirts of \omc{}, and even fewer when adopting the recommended quality cuts, leaving the inner regions ($<5'$) mostly unexplored.

% Opportunity we have in this paper
The oMEGACat survey\footnote{\url{https://omegacatalog.github.io/}} overcomes these limitations by using the Multi Unit Spectroscopic Explorer \citep[MUSE;][]{Bacon2010SPIE}, an integral-field spectrograph that can resolve individual stars within the crowded regions.
This dataset provides over 300,000 MUSE stellar spectra within the half-light radius ($\sim5'$) of \omc{} \citep{Nitschai2023ApJ, Nitschai2024ApJ}, with the wavelength covering 4750-9350~\AA{} at moderate spectral resolution ($R\sim3000$).
With the development of machine learning technique for measuring chemical abundances from MUSE spectra \citep{Wang2022MNRAS}, and its successful demonstration on a globular cluster \citep{Asa'd2024AJ} and the inner Galaxy \citep{Wang2022MNRAS, Wang2025arXiv}, we can now significantly expand the abundance measurements of \omc{} into its most crowded central regions, where previous fiber-fed spectroscopic studies were limited.
By combining this large abundance sample with oMEGACat multi-band photometry \citep{Haberle2024ApJ}, we can build the most comprehensive picture to date of abundance variations across \omc{}'s multiple stellar populations.

% This study
In this paper, we apply the neural network model \textsc{Data-Driven Payne} \citep[hereafter DD-PAYNE;][]{Ting2017ApJL, Xiang2019ApJS} to measure the chemical abundances (C, N, O, Na, Mg, Ca, Ba) from MUSE spectra of \omc{}.
% \textcolor{red}{The main advance of this work is to extend large-sample abundance measurements into the crowded inner regions of \omc{}, where previous fiber-fed spectroscopic studies were limited, and where uniform HST photometry is available for identifying populations.}
We then combine these measurements with literature abundance studies to investigate the abundance and age distributions of \omc{}'s multiple populations.
We describe the oMEGACat dataset, selection criteria, and abundance measurement procedures in Section~\ref{sec:data}.
In Section~\ref{sec:validation}, we validate the reliability of the \ddpayne{} abundance measurements and compare them with high-resolution spectroscopic observations from the literature. 
We also outline the procedures to calibrate the \ddpayne{} abundances to the literature scale.
Section~\ref{sec:results} presents the abundance-metallicity relations and the abundance variations on the chromosome diagram.
In Section~\ref{sec:disscuss}, we examine the abundance differences and their relationships with metallicity, age, and subpopulation, including both the broader P1, Im, and P2 populations \citep{Clontz2025ApJ}, and the more detailed subpopulations in \citet{Clontz2026ApJ}. We then combine our results and discuss potential formation scenarios of different populations in \omc{} in Section~\ref{sec:discuss-senario}.
Finally, a summary of this work and directions for future studies is provided in Section~\ref{sec:summary}.

\section{Data and Methods} 
\label{sec:data}

\subsection{oMEGACat and Quality Cuts} 
\label{subsec:data-oMEGACat}

% General introduction of oMEGACat
We use datasets from the oMEGACat project, which aims to disentangle the dynamics and formation history of \omc{} by compiling the largest spectroscopic, photometric, and astrometric dataset to date.
In brief, the data consists of two main catalogs covering \omc{} within the half-light radius ($\sim5'$). 
The spectroscopic catalog is based on more than 100 MUSE pointings containing a total of 342,797 individual stellar spectra (\citealt{Nitschai2023ApJ}, hereafter \citetalias{Nitschai2023ApJ}). The \textit{HST}-based catalog contains proper motions and seven-band photometry for up to $\sim1.4$ million stars (\citealt{Haberle2024ApJ}, hereafter \citetalias{Haberle2024ApJ})\footnote{The \textit{HST} data were obtained from the Mikulski Archive for Space Telescopes (MAST) at the Space Telescope Science Institute. The specific observations analyzed can be accessed via \dataset[doi:10.17909/26qj-g090]{https://doi.org/10.17909/26qj-g090}.}.
These datasets have already enabled studies of the spatial metallicity distribution \citep{Nitschai2024ApJ}, the age-metallicity relation \citep{Clontz2024ApJ}, helium enrichment \citep{Clontz2025ApJ}, 3D kinematics \citep{Haberle2025ApJ}, the foreground and internal gas kinematics \citep{Wang2025ApJ}, and have resulted in the identification of 14 stellar populations \citep{Clontz2026ApJ}, and the discovery of a central intermediate-mass black hole \citep{Haberle2024Natur}.
In this work, we use the stellar spectra from the oMEGACat MUSE catalog \citepalias{Nitschai2023ApJ} (containing 342,797 spectra) to estimate chemical abundances. 
We refer to \citetalias{Nitschai2023ApJ} for details of the MUSE observations, data reduction, and spectra extraction.

% Selection criteria
We focus on red giant branch (RGB) stars in this study, due to their high signal-to-noise.  
RGB stars are also less affected by atomic diffusion, which can impact the abundance analyses in different populations \citep[e.g.,][]{Dotter2017ApJ}.
Specifically, starting from 342,797 MUSE spectra, we apply the following selection criteria:
\begin{equation}
    \left\{\begin{array}{l}
        P_M>0.95 \\
        m_{\rm F625W}<17~\rm{mag} \\
        m_{\rm F435W}-m_{\rm F625W}>1~\rm{mag} \\
        \rm{qflag_{PampelMUSE}=0} \\
        \rm{S/N}_{\rm{MUSE}}>50~\text{pix}^{-1},
    \end{array}\right.
\label{eqn:quality-cutoff}
\end{equation}
where $P_M$ is the membership probability of \omc{} stars obtained from \citetalias{Nitschai2023ApJ}; we use the threshold recommended in that study, which removes 4,265 stars.
The magnitude and color cuts (both filters from \citealt{Anderson2010ApJ}) then remove 325,691 stars, selecting 12,841 initial RGB candidates.
The PampelMUSE \citep{Kamann2013A&A} quality flag $\rm{qflag_{PampelMUSE}}$ further removes 638 spectra with contributions from multiple sources or source centroids located outside the data cube's field of view.
Finally, the \snrmuse{} threshold removes 4,857 low-S/N spectra, leaving 7,346 stars as the high-quality RGB sample.
We adopt \snrmuse{}~$>50~\rm{pix}^{-1}$ to match the S/N threshold used for the \ddpayne{} training set in \citetalias{Xiang2019ApJS}; below this value, the inferred abundances may be much less reliable.
Our final sample is a subsample of 7,302 stars after the \ddpayne{} fitting quality cutoff, as described in Section~\ref{subsec:data-abundances}.
A color-magnitude diagram (CMD) of the final sample is shown in the left panel of Fig.~\ref{fig:CMD_spectra}.

\subsection{Measuring Chemical Abundances using \ddpayne{}} 
\label{subsec:data-abundances}

\begin{figure*}[!ht]
    \centering
    \includegraphics[width=1\linewidth]{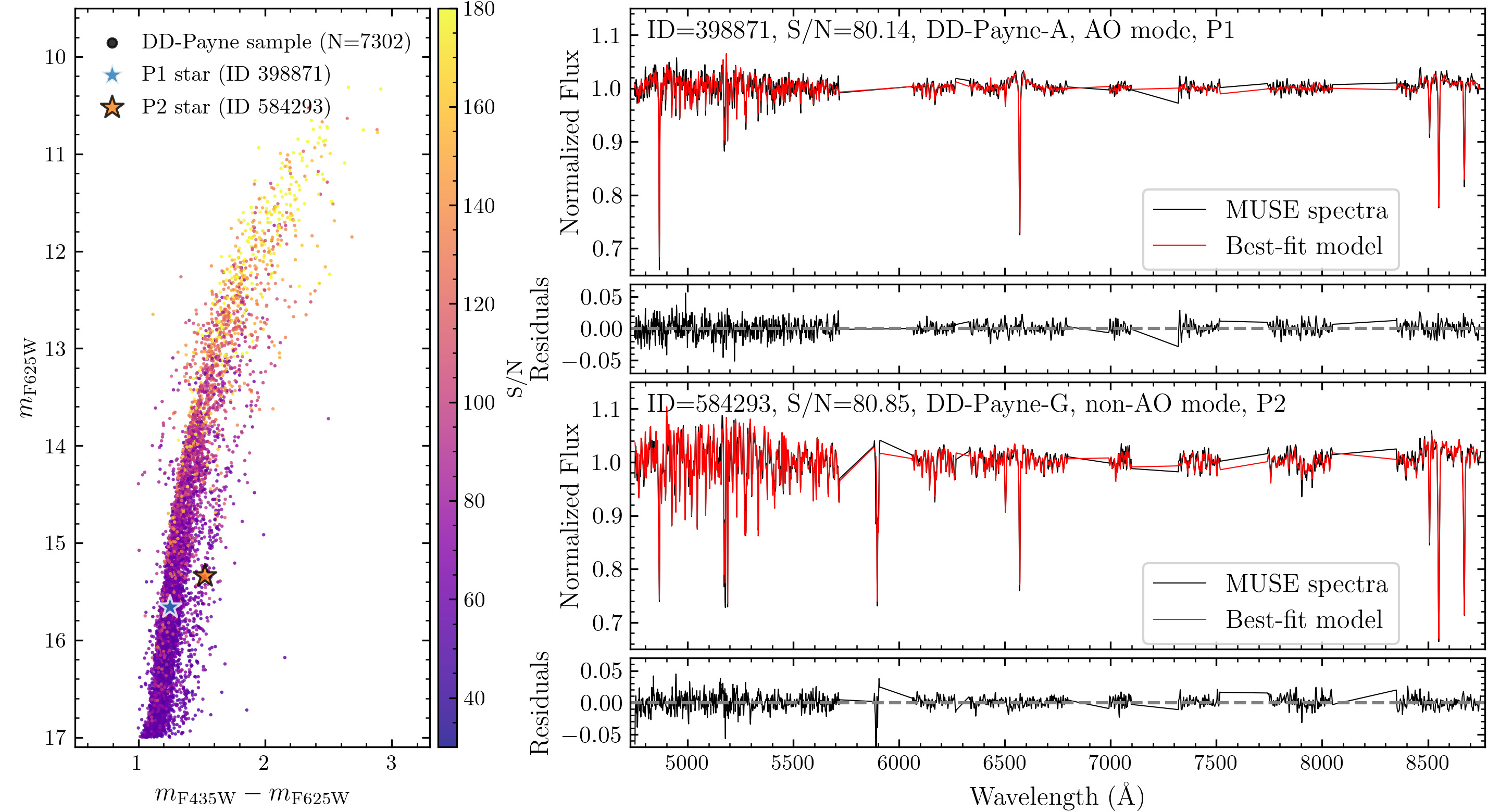}
    \caption{
    \textbf{Left:} Colour-magnitude diagram (CMD) of the oMEGACat \ddpayne{} sample using photometry from \cite{Anderson2010ApJ}, with stars color-coded by the MUSE spectral signal-to-noise ratio (S/N). The two stars shown in the right-hand panels are highlighted as filled star markers. 
    \textbf{Right:} Example \ddpayne{} spectral fits for two stars. We show a primordial (P1) star and a chemically enhanced (P2) star (as defined in Section~\ref{subsec:discuss-stream}), fitted with the \ddpaynea{} and \ddpayneg{} models in AO and non-AO observations, respectively. Note that all stars, in both observing modes, are fitted with all \ddpayne{} models; the cases shown here are representative examples of the fitting performance.
    The spectra are shown in separate wavelength segments because masked wavelength intervals are excluded during the \ddpayne{} fitting, including telluric regions and others affected by LAMOST instrumental interference (see details in \citetalias{Xiang2019ApJS} and \citealt{Wang2022MNRAS}).
    }
    \label{fig:CMD_spectra}
\end{figure*}

% Introduction of DD-Payne
We apply the machine learning model \ddpayne{} \citep[][]{Ting2017ApJL, Xiang2019ApJS} to fit the MUSE spectra and measure chemical abundances.
\ddpayne{} adopts the neural-network interpolator and fitting strategy of \thepayne{} \citep{Ting2019ApJ} combined with physical gradient spectra from Kurucz spectral models \citep{Kurucz1970SAOSR, Kurucz1993KurCD, Kurucz2005MSAIS} to regularize the training process.
The neural network model is trained on LAMOST spectra \citep{Zhao2012RAA, Deng2012RAA, Liu2014IAUS, Liu2015RAA} with stellar labels from high-resolution surveys, including GALAH DR2 \citep{Buder2018MNRAS} and APOGEE-\textsc{Payne} \citep{Ting2019ApJ}, and can derive \teff{}, \logg{}, micro-turbulence velocity ($V_{\rm mic}$), and 16 elemental abundances (\feh{} and \xfe{} of C, N, O, Na, Mg, Al, Si, Ca, Ti, Cr, Mn, Co, Ni, Cu, and Ba) from LAMOST spectra \citep{Xiang2019ApJS}.
A key advantage of \ddpayne{} is that it can enforce these stellar labels (\teff{}, \logg{}, $V_{\rm mic}$, and abundances) to be measured physically from the spectral features, enabling its application to spectra from other instruments.
We refer to \cite{Xiang2019ApJS} (hereafter \citetalias{Xiang2019ApJS}) for more detailed descriptions of the training process and validation of \ddpayne{}. 
\citet{Wang2022MNRAS} applied \ddpayne{} to MUSE spectra and showed that several abundances (Fe, C, Mg, Si, Ti, Ni, Cr) can be measured with accuracies better than 0.1~dex. 
These results demonstrate that, despite line blending at low spectral resolution, models like \ddpayne{} can still provide reliable abundance measurements.

% Procedures to applying DD-Payne
To measure chemical abundances from MUSE spectra of \omc{}, we follow the procedures described by \cite{Wang2022MNRAS}. 
Briefly, we take the spectra from the oMEGACat MUSE observations (\citetalias{Nitschai2023ApJ}, $R\sim3000$) and degrade the spectral resolution to LAMOST ($R\sim1800$).
Then we apply \ddpayne{} to fit the spectra in the wavelength region of 4750-8750~\AA{} using the \texttt{scipy.curve\_fit} routine, and the abundances from the best-fit spectra are adopted as the final measurements.
Here the resolution degradation is required because the \ddpayne{} models were trained on LAMOST spectra, so the spectral fitting should be performed at the LAMOST spectral resolution ($R\sim1800$).
Although this step removes some information relative to the native MUSE resolution, \cite{Wang2022MNRAS} and Section~\ref{sec:validation} here show that the relevant abundance trends can still be recovered statistically from the degraded spectra.
During the fitting, the line-of-sight velocity (\vlos{}) is estimated simultaneously by the Doppler shift equation, with initial \vlos{} set to the values from \citetalias{Nitschai2023ApJ}.
We show two examples of the spectral fitting in the right panels of Fig.~\ref{fig:CMD_spectra}, where the observed MUSE spectra are compared with the \ddpayne{} best-fit for a P1 star (AO mode) and a P2 star (non-AO mode).
In total, \ddpayne{} provides \teff{}, \logg{}, and 16 chemical abundances. 
Here we only focus on [C/Fe], [N/Fe], [O/Fe], [Na/Fe], [Mg/Fe], [Ca/Fe], and [Ba/Fe], which are well-known elements that show correlations or variations in different populations of globular clusters \citep[e.g.,][]{Johnson2010ApJ, Gratton2012A&ARv, Bastian2018ARA&A}.
\citet{Wang2022MNRAS} reported typical precisions better than 0.2~dex for these elements from MUSE (see their Section~4 for details).
It is worth noting that although \ddpayne{} provides [Al/Fe], the associated uncertainty is large because most Al features lie outside the MUSE wavelength range (see Fig.~22 of \citetalias{Xiang2019ApJS}).
Therefore, we do not analyze [Al/Fe] from \ddpayne{} in this study.
However, we include [Ba/Fe] because several strong absorptions can be detected by \ddpayne{} (see details in Section~\ref{subsec:validation-ddp}), and that \citetalias{Xiang2019ApJS} found it can be reasonably measured down to $\feh{}=-4$~dex.
This \textit{s}-process element can also provide useful clues on the formation history of \omc{} \citep[e.g., ][]{Marino2011ApJ}.

% Selection of chemical abundances from which model
\ddpayne{} has two neural network models, \ddpaynea{} (APOGEE-based) and \ddpayneg{} (GALAH-based), and both can provide reliable abundance measurements.
Although \citetalias{Xiang2019ApJS} provided recommendations for each element based on their validation tests, we preferentially adopt \ddpaynea{} where available and use \ddpayneg{} for abundances not provided by \ddpaynea{} (notably [Na/Fe] and [Ba/Fe]).
This is because almost all of our \omc{} stars have metallicity lower than $-0.5$~dex, while the \ddpayneg{} was trained primarily on stars with $\feh{}>-0.8$~dex.
Fig.~6 of \citetalias{Xiang2019ApJS} shows that \ddpayneg{} becomes more uncertain in \teff{}, \logg{}, and \feh{} at low metallicity, while \ddpaynea{} remains robust down to $\feh{}\approx-1.5$~dex.
Therefore, we adopt \teff{}, \logg{}, \feh{}, [C/Fe], [N/Fe], [O/Fe], [Mg/Fe], [Ca/Fe] from \ddpaynea{} and [Na/Fe] and [Ba/Fe] from \ddpayneg{}.
We note that because [Na/Fe] and [Ba/Fe] are taken from \ddpayneg{} in a much lower metallicity regime than its training set, these abundances are measured purely from extrapolation and therefore require additional caution.
We refer to Section~\ref{sec:validation} for detailed validations of all the abundances.

From the MUSE spectra of 7,346 oMEGACat RGB stars, we apply the following selection criteria to select stars with good \ddpayne{} fitting:
\begin{equation}
    \left\{\begin{array}{l}
        \text{red}\_\chi^2_{\ddpaynea{}}<20 \\
        \text{red}\_\chi^2_{\ddpayneg{}}<20 \\
        \feh{}>-2.394~\rm{dex}, \\
    \end{array}\right.
\label{eqn:ddp-cutoff}
\end{equation}
where $\text{red}\_\chi^2_{\ddpaynea{}}$ and $\text{red}\_\chi^2_{\ddpayneg{}}$ are the reduced chi-square values from the \ddpayne{} spectral fitting.
We adopt $\text{red}\_\chi^2<20$ according to the LAMOST-MUSE cross-validation in \citet{Wang2022MNRAS}, where the reduced-$\chi^2$ values of common validation stars are mostly lower than 25.
This threshold removes 41 stars with poorly fitted spectra.
We also tested thresholds of $\text{red}\_\chi^2<10$ and $\text{red}\_\chi^2<25$ and found no significant change on the final results of this work.
The metallicity cut (using the uncalibrated \feh{}) at $-2.394$~dex excludes 3 stars whose inferred metallicities lie at the edge of the model's fitting range, where the \ddpayne{} measurements become unreliable.
After applying these criteria, 7,302 stars remain in the sample, which we refer to as the oMEGACat \ddpayne{} catalog.

\section{Validation of DD-Payne Chemical Abundances} 
\label{sec:validation}

Before proceeding to the scientific analysis, we validate the chemical abundances derived with \ddpayne{}. 
We first examine whether \ddpayne{} correctly identifies the relevant spectral features for each element, particularly in the low-metallicity regime. 
We then directly compare the inferred abundances with high-resolution spectroscopic measurements from the literature using cross-matched samples. 
Finally, we calibrate the \ddpayne{} abundances onto a consistent literature scale to mitigate systematic offsets.

% \begin{itemize}
%     \item We first select DDPayne abundances following recommendations fron Xiang et al. 2019. This is based on their validation process
%     \item We then calculate the gradient correlation at \feh{} at -2.5, -1.5, -0.5, to access their precision. If the correlation is low, we won't use it because DDPayne can not pick up the correct absorption feature change from MUSE spectra.
%     \item We then compare from literature using crossmatched sample, calculate median and std of abundances difference, and its relation with metallicity
%     \item We then apply metallicity-dependent correction to the abundances because although DDPayne correctly learns the absorption features, these stars are still estimated by extrapolation. Therefore, there could be some offset issue.
%     \item However, we still include \ofe{} when studying Na-O relation, because with comparison to Johnson et al. 2010, this relation is pretty reliable.
%     \item Conclusion: C, N, O only for [Fe/H]>-1. Na, Ba for all. Mg is with [Fe/H]>-1.5
% \end{itemize}

\subsection{Measurement Robustness of \ddpayne{}}
\label{subsec:validation-ddp}

\begin{deluxetable}{lccc}
\tabletypesize{\footnotesize}
\tablewidth{0pt}
\tablecaption{Pearson correlation coefficients ($r$) between \ddpayne{} gradient spectra (from \citetalias{Xiang2019ApJS}) and Kurucz models in the LAMOST spectral resolution ($R\sim1800$) and MUSE fitting wavelength range (4750-8750~\AA{}) at three metallicities ($\feh{}=-2.5$, $-1.5$, and $-0.5$~dex).}
\label{tab:ddpayne_kurucz_grad_corr}
\tablehead{
\colhead{Parameter} & \colhead{$\feh{}=-2.5$} & \colhead{$\feh{}=-1.5$} & \colhead{$\feh{}=-0.5$}
}
\startdata
\teff{}$^{a}$    & 0.768 & 0.806 & 0.785 \\
\logg{}$^{a}$    & 0.426 & 0.568 & 0.656 \\
{\feh{}}$^{a}$   & 0.866 & 0.896 & 0.849 \\
{[C/Fe]}$^{a}$   & 0.124 & 0.207 & 0.782 \\
{[N/Fe]}$^{a}$   & 0.018 & 0.117 & 0.681 \\
{[O/Fe]}$^{a}$   & 0.021 & 0.056 & 0.621 \\
{[Na/Fe]}$^{b}$  & 0.987 & 0.996 & 0.997 \\
{[Mg/Fe]}$^{a}$  & 0.209 & 0.476 & 0.716 \\
{[Ca/Fe]}$^{a}$  & 0.701 & 0.746 & 0.736 \\
{[Ba/Fe]}$^{b}$  & 0.896 & 0.941 & 0.989 \\
\enddata
\tablecomments{
(a) Gradient spectra from \ddpaynea{}; (b) Gradient spectra from \ddpayneg{}.}
\end{deluxetable}

% Introduce gradient spectra
As noted by \citetalias{Xiang2019ApJS}, the key factor enabling \ddpayne{} to recover chemical abundances reliably is the regularization of its training process by \abinitio{} Kurucz spectral models \citep{Kurucz1970SAOSR, Kurucz1993KurCD, Kurucz2005MSAIS}.
During the training, \ddpayne{} calculates the correlation coefficients ($r$) between its own gradient spectra and those from Kurucz models.
The gradient spectrum is defined as $\delta f_l / \delta l$, where $l$ is the stellar parameter and $f_l$ is the predicted flux.
These gradients reflect how the flux changes with each parameter, allowing the model to be constrained toward physically meaningful spectral responses even when the training set is limited.
High correlations ($r\ge0.5$) indicate that \ddpayne{} can accurately identify the absorption features of each element, and low correlations ($r<0.5$) indicate that the abundance measurements from \ddpayne{} are purely from astrophysical correlations of the training sets.
Therefore, before analyzing the \omc{} MUSE spectral abundances, it is essential to confirm whether \ddpayne{} correctly identifies their spectral features.

% Calculate the gradient correlation at \feh{} at -2.5, -1.5, -0.5, to access their precision.
To do this, we take three of the 16 sets of Kurucz gradient spectra calculated by \citetalias{Xiang2019ApJS} (see details in their Table~1) with \feh{} of $-2.5$, $-1.5$ and $-0.5$, scaled-solar abundance ratios ([X/Fe]=0), and $\logg{}<2.5$~dex.
We choose these three sets because their \feh{} and \logg{} are the closest available matches to our \omc{} RGB stars.
We then compute the Pearson correlation coefficients ($r$) between these Kurucz gradient spectra and those predicted by \ddpayne{} with the same stellar parameters.

The results are listed in Table~\ref{tab:ddpayne_kurucz_grad_corr}.
Overall, the correlations show the expected metallicity dependence, where the ability of \ddpayne{} to recover abundances declines as spectral features weaken at lower metallicities.  
For \teff{}, \logg{}, and \feh{}, the coefficients are high across all metallicities, confirming that they can be reliably measured.  
For [C/Fe], [N/Fe], and [O/Fe], the correlations are low at $\feh{}=-2.5$ and $-1.5$~dex, and the values become larger than 0.5 only at $\feh{}=-0.5$~dex.  
This indicates that at very low metallicity, the relevant features become too weak for \ddpayne{} to identify. 
The [Mg/Fe] coefficients increase with metallicity, and the value becomes $\sim$0.5 when $\feh{}>-1.5$~dex.  
The [Ca/Fe] coefficients are higher than 0.5 for all three metallicities, mostly due to the CaT absorptions (also shown in Fig.~\ref{fig:CMD_spectra}).
For the abundances derived from the GALAH-based model (\ddpayneg{}), both [Na/Fe] and [Ba/Fe] show correlation coefficients close to 1 at all metallicities.  
This indicates spectral features of these two abundances can be correctly identified.
However, because most stars in the \ddpayneg{} training set have $\feh{}>-0.8$~dex, these high correlations at lower metallicity are driven primarily by the Kurucz model regularization rather than by the training data.
Thus, although the spectral features of [Na/Fe] and [Ba/Fe] are physically constrained by \ddpayneg{}, their absolute abundance values may still be biased. 
They should therefore be tested and, if necessary, calibrated against external high-resolution spectroscopic measurements, as we do for these and the other abundances in Section~\ref{subsec:validation-direct}.

% Conclusion of gradient spectra correlations
% Overall, the correlations in Table~\ref{tab:ddpayne_kurucz_grad_corr} show the expected metallicity dependence, where the ability of \ddpayne{} to recover abundances declines as spectral features weaken in lower metallicities.  
% We therefore conclude that the \ddpaynea{} model provides physically meaningful results for \teff{}, \logg{}, \feh{} for all metallicities and [C/Fe], [N/Fe], and [O/Fe] when $\feh{}>-1.0$~dex, whereas [Mg/Fe] measurements are physical when $\feh{}>-1.5$~dex.
% [Na/Fe] and [Ba/Fe] from \ddpayneg{} remain usable across the full metallicity range, although with potential systematic offsets. 
Based on these correlations, in this study, we adopt \ddpaynea{} measurements of \teff{}, \logg{}, \feh{}, and [Ca/Fe] at all metallicities, [C/Fe], [N/Fe], and [O/Fe] for $\feh{}>-1.0$~dex, and [Mg/Fe] for $\feh{}>-1.5$~dex. 
For [Na/Fe] and [Ba/Fe], we use the \ddpayneg{} measurements across the full metallicity range.
In Section~\ref{subsec:validation-direct}, we examine the abundance measurement uncertainties through direct comparisons with high-resolution observations from the literature.
In Section~\ref{subsec:validation-calibration}, we use these comparisons to calibrate any systematic offsets before using the abundances for scientific analysis.
We note that the calibration robustness varies between elements because the cross-matched literature samples differ in size, metallicity coverage, and abundance-analysis methods.
Therefore, the calibration is intended to remove the dominant systematic offsets and place the \ddpayne{} measurements on an approximate literature scale, rather than to provide a perfect absolute calibration for individual stars.
% All adopted abundances are subsequently calibrated to literature values from high-resolution spectroscopic studies to place them on a consistent abundance scale and reduce systematic offsets.

\subsection{Comparing \ddpayne{} Abundances to Literature using Cross-matched Samples}
\label{subsec:validation-direct}

% Outline
To further test the accuracy of our \ddpayne{} abundances, in this section, we directly compare our abundance measurements with literature results derived from high-resolution spectroscopic observations of \omc{}.
This comparison provides a direct validation of both the absolute abundance scale and the abundance trends recovered from MUSE spectra.

\subsubsection{Cross-matching with Literature}
\label{subsubsec:validation-direct-crossmatch}

\begin{deluxetable*}{l*{5}{ccc}}
\label{tab:literature_counts}
\tablecaption{Residual statistics in the cross-matched sample between the oMEGACat \ddpayne{} catalog and literature datasets. For each quantity we list the number of matched stars ($N$), mean residual $\mu(\Delta)$, and scatter $\sigma(\Delta)$, where $\Delta X \equiv X_{\rm \ddpayne}-X_{\rm lit}$.}
\tablehead{
\colhead{Quantity} &
\multicolumn{3}{c}{Johnson10+} &
\multicolumn{3}{c}{Marino11/12+} &
\multicolumn{3}{c}{M\'esz\'aros21+} &
\multicolumn{3}{c}{AlvarezGaray22/24+} &
\multicolumn{3}{c}{Schiavon24+} \\
\cline{2-4}\cline{5-7}\cline{8-10}\cline{11-13}\cline{14-16}
& \colhead{$N$} & \colhead{$\mu(\Delta)$} & \colhead{$\sigma(\Delta)$}
& \colhead{$N$} & \colhead{$\mu(\Delta)$} & \colhead{$\sigma(\Delta)$}
& \colhead{$N$} & \colhead{$\mu(\Delta)$} & \colhead{$\sigma(\Delta)$}
& \colhead{$N$} & \colhead{$\mu(\Delta)$} & \colhead{$\sigma(\Delta)$}
& \colhead{$N$} & \colhead{$\mu(\Delta)$} & \colhead{$\sigma(\Delta)$}
}
\startdata
$T_{\rm eff}$~\rm{(K)}     & 378 & -96.20 &  66.88 & 172 & 293.65 &  93.46 &  66 &  69.68 & 113.46 & 223 & 100.56 & 130.22 & 195 & 129.83 &  95.34 \\
$\log g$                  & 378 &   0.61 &   0.22 & 172 &   0.42 &   0.24 &  66 &   0.27 &   0.26 & 223 &   0.36 &   0.23 & 195 &   0.39 &   0.23 \\
$\mathrm{[Fe/H]}$         & 378 &   0.32 &   0.12 & 172 &   0.36 &   0.11 &  66 &   0.11 &   0.13 & 223 &   0.31 &   0.09 & 195 &   0.23 &   0.07 \\
$\mathrm{[C/Fe]}$         &   0 & \nodata & \nodata &  70 &   0.37 &   0.46 &  33 &  -0.14 &   0.51 &   0 & \nodata & \nodata & 195 &   0.16 &   0.21 \\
$\mathrm{[N/Fe]}$         &   0 & \nodata & \nodata &  70 &  -0.40 &   0.43 &  12 &  -0.53 &   0.35 &   0 & \nodata & \nodata & 195 &  -0.39 &   0.31 \\
$\mathrm{[O/Fe]}$         & 376 &   0.33 &   0.29 & 153 &   0.16 &   0.38 &  23 &  -0.20 &   0.25 &   0 & \nodata & \nodata & 194 &   0.33 &   0.41 \\
$\mathrm{[Na/Fe]}$        & 299 &   0.23 &   0.28 & 142 &   0.18 &   0.29 &   3 &   0.36 &   0.18 & 148 &   0.31 &   0.26 & 143 &   0.30 &   0.52 \\
$\mathrm{[Mg/Fe]}$        &   0 & \nodata & \nodata &   0 & \nodata & \nodata &  60 &  -0.13 &   0.15 & 186 &  -0.09 &   0.18 & 195 &   0.05 &   0.15 \\
$\mathrm{[Ca/Fe]}$        & 378 &  -0.09 &   0.10 &   0 & \nodata & \nodata &  33 &  -0.13 &   0.15 &   0 & \nodata & \nodata & 191 &  -0.01 &   0.12 \\
$\mathrm{[Ba/Fe]}$        &   0 & \nodata & \nodata & 172 &   0.96 &   0.48 &   0 & \nodata & \nodata &   0 & \nodata & \nodata &   0 & \nodata & \nodata \\
\enddata
\end{deluxetable*}

We select five literature works that each have several hundred high-resolution spectroscopic observations of RGB stars in \omc{}. 
The characteristics of these studies are summarized below:
\begin{itemize}
    \item \cite{Johnson2010ApJ}: 855 RGB stars observed with Blanco/Hydra ($R\sim18,000$), providing abundances of Fe, O, Na, Al, Si, Ca, Sc, Ti, Ni, La, and Eu.
    \item \cite{Marino2011ApJ, Marino2012ApJ}: 300 RGB stars observed with VLT/FLAMES ($R\sim20,000$-$25,000$), measuring elements of Fe, Na, C, N, O, Ba, and La, where the C and N measurements are from \cite{Marino2012ApJ} for 77 stars.
    \item \cite{Meszaros2021MNRAS}: 982 RGB stars observed with APOGEE DR16 ($R\sim22,500$), including elements of Fe, C, N, O, Mg, Al, Si, K, Ca, and Ce measured using the BACCHUS code.
    \item \cite{AlvarezGaray2022ApJL, AlvarezGaray2024AA}: 450 RGB stars observed with VLT/FLAMES ($R\sim20,000$-$29,000$), providing elements of Fe, Na, Mg, Al, Si, and K.
    \item \cite{Schiavon2024MNRAS}: 1,864 giant stars from APOGEE DR 17 with up to 20 elemental abundances measured from the ASPCAP pipeline.
\end{itemize}

% Crossmatch strategy
% Following the RGB selection defined in Equation~\ref{eqn:quality-cutoff}, 
% First, we applied a magnitude cut in $m_{\rm F625W}$ (from \citealt{Anderson2010ApJ}) to restrict our oMEGACat sample to the RGB region to reduce contamination by fainter stars.  
We perform positional cross-matching between the oMEGACat \ddpayne{} catalog (obtained in Section~\ref{subsec:data-abundances}) and each of these studies using right ascension ($\alpha$) and declination ($\delta$), and apply positional separation criteria to be less than 1~arcsec.  
The number of successfully matched stars with available abundances is listed in Table~\ref{tab:literature_counts}. 
Because our oMEGACat \ddpayne{} dataset focuses on \omc{} central regions and most literature stars are at larger radii, only a small fraction of stars are cross-matched.
Nevertheless, the cross-matched samples with more than 100 stars are sufficient for a direct comparison of their abundance measurements.

\subsubsection{Comparing the Chemical Abundances}
\label{subsubsec:validation-direct-compare}

\begin{figure*}[!ht]
    \centering
    \includegraphics[width=1\linewidth]{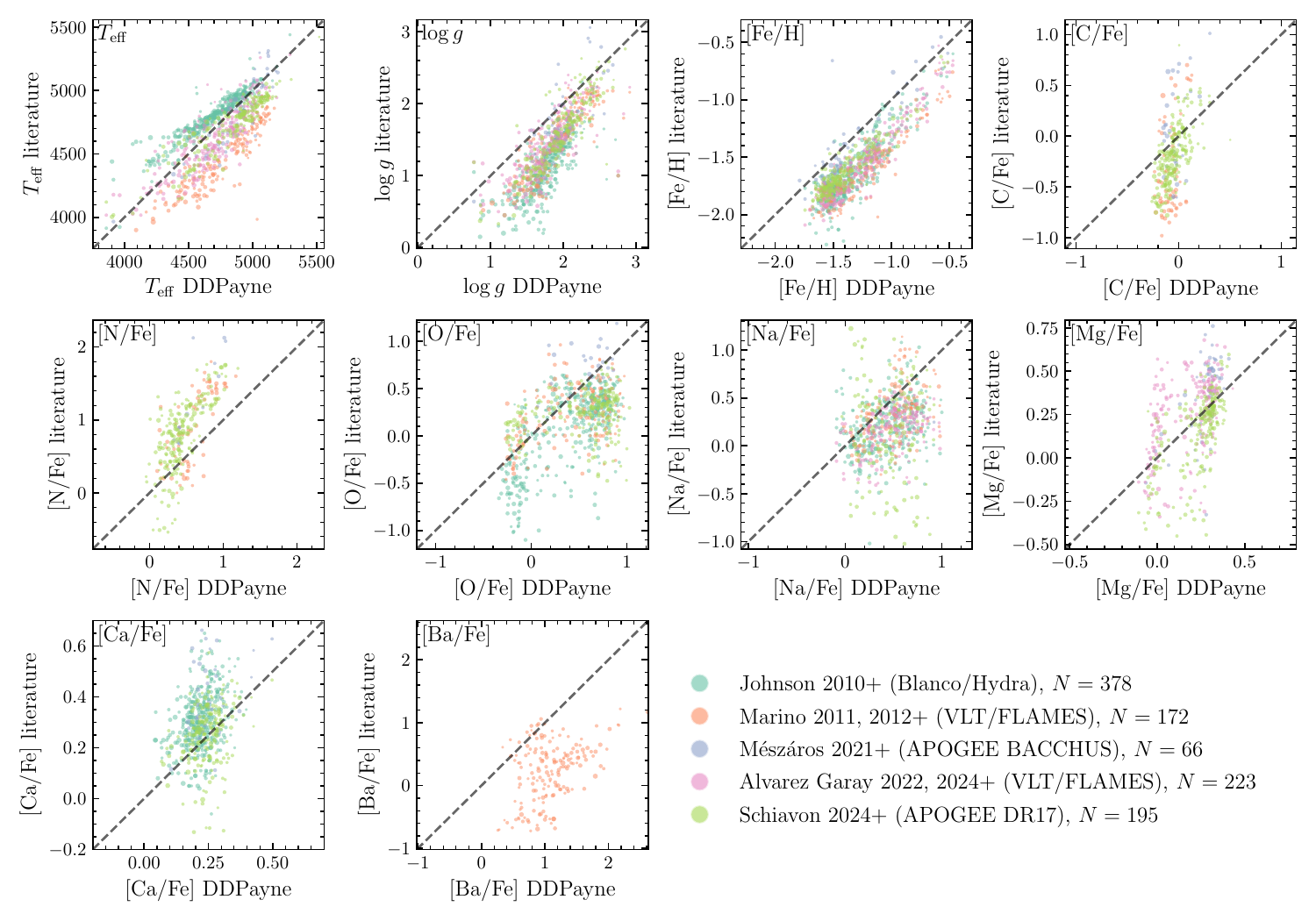}
    \caption{
    Direct comparison of \ddpayne{} abundance measurements for oMEGACat \ddpayne{} stars with literature values for the cross-matched samples.
    Common stars between oMEGACat and each literature catalog are shown in different colors, with point size indicating \snrmuse{}.
    The number of common stars is provided in the legend for each literature work, and the black dashed lines are the one-to-one relation for visual reference.
    }
    \label{fig:abundances_1_1_literature}
\end{figure*}

% Figure. Direct comparison
Fig.~\ref{fig:abundances_1_1_literature} presents one-to-one comparisons of our \ddpayne{} abundance measurements with literature values for the cross-matched stars.  
Each color corresponds to a literature study, and the point sizes correlate with \snrmuse{}. 
The black dashed lines indicate the one-to-one relation.
Note that we do not show the formal uncertainties of each stellar parameter from the \ddpayne{} spectral fitting in Fig.~\ref{fig:abundances_1_1_literature} given studies have shown that they are underestimated \citep{Wang2022MNRAS, Sandford2020ApJS}. Instead, we use their deviations from literature measurements as an indicator of the measurement uncertainties.
% throughout this work, assuming the literature measurements are relatively more robust.}
As shown in Fig.~\ref{fig:abundances_1_1_literature}, \teff{}, \logg{}, and \feh{} show strong correlations with literature values but with a constant offset.
This confirms that \ddpayne{} provides reliable measurements of \teff{}, \logg{}, and \feh{} from MUSE spectra.  
For [C/Fe] and [N/Fe], the literature measurements span a wider abundance range, while the \ddpayne{} results show narrower distributions. 
[O/Fe] and [Na/Fe] demonstrate larger scatters than \teff{}, \logg{}, and \feh{}.  
The [Mg/Fe] comparison shows two sequences, which is likely because the \ddpaynea{} training set is dominated by Milky Way field stars and the model tends to reproduce the $\alpha$-bimodality of the Galactic disk \citep[e.g.,][]{Hayden2015ApJ}. However, the literature measurements also exhibit a similar but less pronounced bimodal feature (see discussions in Section~\ref{subsec:results-abundances-feh}), indicating that this structure might not be solely driven by the training data.
We will assess the impact on Mg measurements in detail in Section~\ref{subsec:results-abundances-feh} and \ref{subsec:discuss-stream}.
Another $\alpha$-element [Ca/Fe] also shows a scattered distribution, but the ranges on both axes are narrower than for the other elements.
The $s$-process element [Ba/Fe] shows a constant offset, with significant scatter at [Ba/Fe]~$\sim1$~dex. % aligning well to the one-to-one relation.  
This figure quantitatively demonstrates the difference in chemical abundances measured from \ddpayne{} and high-resolution spectroscopic observations in the literature.

\begin{figure}
    \centering
    \includegraphics[width=1\linewidth]{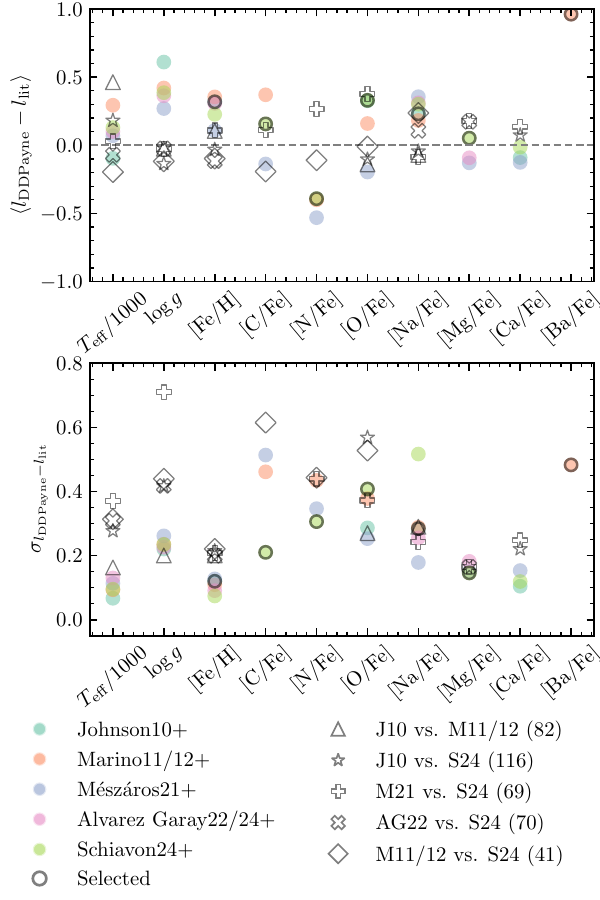}
    \caption{
    Median (top panel) and standard deviation (bottom panel) of the differences between oMEGACat \ddpayne{} abundances and literature studies for the cross-matched stars.
    The color scheme for each literature is the same as in Fig.~\ref{fig:abundances_1_1_literature}.
    Gray markers with different shapes indicate comparisons of cross-matched stars between different literature, with the number of stars shown in parentheses.
    The open circle marker labeled ``Selected'' indicates the reference literature used for the abundance calibration of each element. 
    As discussed in detail in Section~\ref{subsec:validation-calibration}, we adopt \feh{}, and [Na/Fe] from \cite{Johnson2010ApJ}; [C/Fe], [O/Fe], [N/Fe], and [Mg/Fe] from \cite{Schiavon2024MNRAS}; and [Ba/Fe] from \cite{Marino2011ApJ} as references to calibrate \ddpayne{} abundances.
    }
    \label{fig:abundances_precision_literature}
\end{figure}

% Figure. Precision and literature comparison
Although Fig.~\ref{fig:abundances_1_1_literature} shows scatters in the direct abundance comparison between \ddpayne{} and literature, one thing to note is that the literature measurements are not the ``true'' answers as they also have measurement errors.
To further quantify the abundance differences, we plot in Fig.~\ref{fig:abundances_precision_literature} the median and standard deviation of the \ddpayne{} abundance difference relative to the literature measurements, with the values also shown in Table~\ref{tab:literature_counts}.
We also plot in gray the median and dispersion between different literature studies themselves using their cross-matched stars for comparison.
As Fig.~\ref{fig:abundances_precision_literature} shows, the abundance difference between literature studies is often comparable to that between \ddpayne{} and literature.
Notably, [Na/Fe] from \citep{Schiavon2024MNRAS}  (bottom panel) shows a much larger scatter than the other elements due to the highly uncertain Na measurements from APOGEE spectra, as discussed in \cite{Shetrone2026ApJ}.
Overall, this figure implies that \ddpayne{}-inferred abundance uncertainties are of the same order as inter-study systematics in high-resolution spectroscopic observations.
Therefore, although \ddpayne{} does not reach the abundance precision of high-resolution spectroscopy for individual stars, given the larger sample, its measurements are sufficiently robust for statistical analyses of abundance relations.
Once the \ddpayne{}-inferred abundances are calibrated to the high-resolution scale, the dataset can be combined with existing literature to significantly expand the abundance sample of \omc{}, particularly within its half-light radius and in the relatively metal-rich regime ($\feh{}>-1$~dex). This enables a more comprehensive characterization of \omc{}'s chemical abundance patterns and its enrichment history.

\subsection{Calibrating \ddpayne{} Abundances to the Literature Scale}
\label{subsec:validation-calibration}

\begin{figure*}[!ht]
    \centering
    \includegraphics[width=1\linewidth]{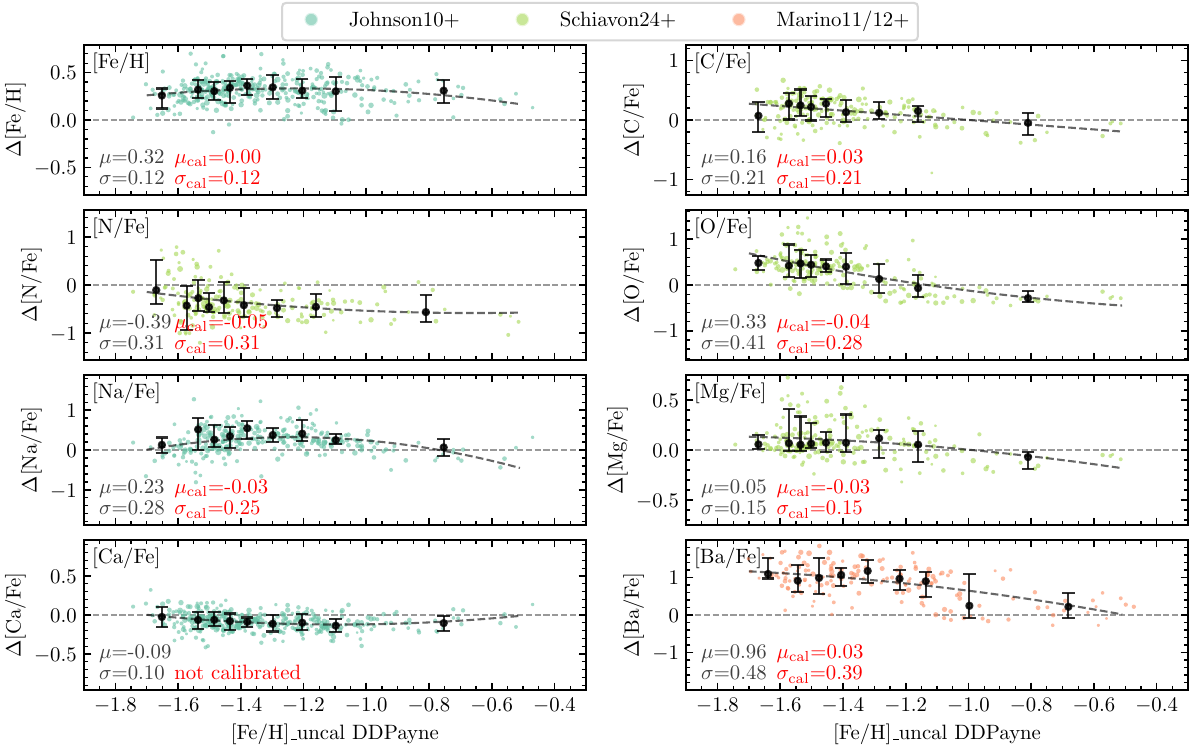}
    \caption{
    Difference between oMEGACat \ddpayne{} abundances and literature measurements ($\Delta$[X/Fe]) as a function of \feh{} (original \ddpayne{} measurements without calibration).
    For each element, we use the standard reference literature adopted in Section~\ref{subsec:validation-calibration}.
    The stars are divided into nine \feh{} bins, and the median $\Delta$[X/Fe] values are plotted as black points with error bars represented by the $16^{th}$ and $84^{th}$ percentiles.
    The gray dashed lines represent the fitted second-degree polynomial relations used to calibrate the \ddpayne{} abundances to the literature scale, as described in Section~\ref{subsec:validation-calibration}.
    In each panel, the black annotations in the lower-left corner report the overall median $\Delta$[X/Fe] and $\sigma\equiv(p_{84}-p_{16})/2$ computed from all stars shown in that panel; these values are also listed in Table~\ref{tab:literature_counts}.
    The red annotations show the same values after applying the metallicity-dependent calibration on the \ddpayne{}-measured abundances as described in Section~\ref{subsec:validation-calibration}.
    }
    \label{fig:abundances_literature_metallicity_selection}
\end{figure*}

% \subsection{Abundances Selection}
% \label{subsec:validation-selection}
% Abundances Selection

% Explaination of why we need abundance calibration
% As discussed in Section~\ref{subsec:validation-direct}, the abundance differences between \ddpayne{} and literature measurements are comparable to the variations between different literature values.
% However, systematic offsets may remain in the absolute values.
% As discussed in Section~\ref{subsec:validation-ddp}, while \ddpayne{} can identify the corresponding spectral absorption features for Na and Ba down to $\feh{}=-2.5$~dex (as shown in Table~\ref{tab:ddpayne_kurucz_grad_corr}), the absolute abundance estimates may still be biased.
% Therefore, it is necessary to apply an abundance calibration to place the \ddpayne{} measurements on the same scale as high-resolution spectroscopic studies.
% Then we can combine \ddpayne{} measurements and the literature as a whole to investigate the chemical abundances of \omc{}.

% Selected literature as the standard for calibration
To calibrate \ddpayne{} abundances to the literature scale, we first determine which literature to adopt as the ``standard'' reference for each element.
The selected references are marked with gray circles in Fig.~\ref{fig:abundances_precision_literature}.
Specifically, we adopt \feh{}, and [Na/Fe] from \cite{Johnson2010ApJ}; [C/Fe], [O/Fe], [N/Fe], and [Mg/Fe] from \cite{Schiavon2024MNRAS}; and [Ba/Fe] from \cite{Marino2011ApJ}.
These choices are either due to the smallest standard deviation of abundance differences relative to \ddpayne{} (bottom panel of Fig.~\ref{fig:abundances_precision_literature}) or the largest number of cross-matched stars, as summarized in Table~\ref{tab:literature_counts}.
Notably, we adopt [C/Fe], [N/Fe], and [O/Fe] from the same literature to avoid systematic biases from different measurement methods, which is essential for investigating the [(C+N+O)/Fe] distribution later in this study.
As for [Ca/Fe], we do not perform an abundance calibration because it shows the smallest deviation ($\sigma(\Delta)\sim0.1$~dex) compared to other parameters.

% Show Delta[X/Fe] vs. [Fe/H] figure
We then use these reference datasets to calibrate the \ddpayne{} abundance measurements.
First, we plot the abundance differences ($\Delta$[X/Fe]) between \ddpayne{} and the reference literature as a function of metallicity, as shown in Fig.~\ref{fig:abundances_literature_metallicity_selection}.
For each element, we show only the comparison with the selected reference study, using the same color scheme as in Fig.~\ref{fig:abundances_1_1_literature}.
We divide the stars into nine \feh{} bins and calculate the median and 16th-84th percentile values (black points with error bars) to evaluate how the offsets vary with metallicity.
As shown in Fig.~\ref{fig:abundances_literature_metallicity_selection}, \feh{} shows nearly constant offsets, whereas other elements show larger median offsets and scatter at lower metallicities.
This behavior is expected because the \ddpayne{} training set includes fewer stars with $\feh{}<-1$~dex.
In particular, $\Delta$[Ba/Fe] shows a clear transition near $\feh{}\sim-1$~dex, changing from positive values in the metal-poor regime to values near zero at higher metallicity.
This explains the [Ba/Fe] behavior we see in Fig.~\ref{fig:abundances_1_1_literature}, where the stars align closest to the one-to-one relation should be mostly metal-rich.
Although we do not apply a calibration to [Ca/Fe], we still compare it with \cite{Johnson2010ApJ} in Fig.~\ref{fig:abundances_literature_metallicity_selection} as a showcase of \ddpayne{} reliability. The resulting $\mu$ and $\sigma$ values show that it is highly consistent with literature.

% Explain procedures of abundance calibration
To correct these systematic offsets, we apply a metallicity-dependent calibration using the $\Delta$[X/Fe]-metallicity relations shown in Fig.~\ref{fig:abundances_literature_metallicity_selection}.
For each element, we fit a second-degree polynomial to model the offset as a function of \feh{} (originally from \ddpayne{} without calibration), and then subtract this calibration term ($\Delta$[X/Fe]) from the original \ddpayne{} abundances.
We adopt a second-order polynomial because it is the lowest-order function that captures the smooth metallicity-dependent offsets in Fig.~\ref{fig:abundances_literature_metallicity_selection}, especially the curvature in [O/Fe], [Na/Fe], and [Ba/Fe].
The fitted polynomial relations are shown as black dashed lines in Fig.~\ref{fig:abundances_literature_metallicity_selection}.
The red annotations show the median residual $(\Delta{\rm [X/Fe]})$ and $\sigma\equiv(p_{84}-p_{16})/2$ after the abundance calibration.
These values provide empirical external uncertainty estimates, including both \ddpayne{} measurement uncertainties and literature systematics, and are more representative than the formal \ddpayne{} fitting uncertainties.
In the following analysis, all abundance measurements are based on these calibrated values, except for [Ca/Fe], unless otherwise noted.
For metallicity, we use $\rm{[Fe/H]_{uncal}}$ to denote the original \ddpayne{} metallicity measurement, and \feh{} to denote the calibrated metallicity.

\section{Results}
\label{sec:results}

\subsection{Abundance Quality Selection of oMEGACat \ddpayne{} Catalog} 
\label{subsec:results-quality}

\begin{deluxetable}{l p{0.6\columnwidth} l}
\tabletypesize{\footnotesize}
\tablecaption{Column descriptions of the oMEGACat \ddpayne{} catalog, published as an online machine-readable table.}
\tablehead{
\multicolumn{1}{l}{Column} &
\multicolumn{1}{l}{Description} &
\multicolumn{1}{l}{Unit}
}
\startdata
R.A.          & Right Ascension (J2000) from \citetalias{Nitschai2023ApJ} & deg \\
Decl.         & Declination (J2000) from \citetalias{Nitschai2023ApJ} & deg \\
ID\_MUSE      & oMEGACat MUSE catalog identifier from \citetalias{Nitschai2023ApJ} &  \\
% ID\_AvdM      & Identifier from \cite{Anderson2010ApJ} &  \\
% ID\_H24       & Identifier from \cite{Haberle2024ApJ} &  \\
ID\_J10       & Identifier from \cite{Johnson2010ApJ} &  \\
ID\_M1112     & Identifier from \cite{Marino2011ApJ, Marino2012ApJ} &  \\
ID\_M21       & Identifier from \cite{Meszaros2021MNRAS} &  \\
ID\_AG2224    & Identifier from \cite{AlvarezGaray2022ApJL, AlvarezGaray2024AA} &  \\
ID\_S24       & Identifier from \cite{Schiavon2024MNRAS} &  \\
S/N           & Average MUSE spectral signal-to-noise ratio on the full wavelength & \perpixel{} \\
\teff{}       & Effective temperature from \ddpayne{} & K \\
e\_\teff{}    & Formal uncertainty in \teff{} & K \\
\logg{}       & Surface gravity from \ddpayne{} & dex \\
e\_\logg{}    & Formal uncertainty in \logg{} & dex \\
$[\mathrm{Fe/H}]$           & Metallicity from \ddpayne{} & dex \\
e\_[Fe/H]     & Formal uncertainty in $[\mathrm{Fe/H}]$ & dex \\
$[\mathrm{X/Fe}]$        & Chemical abundance in [X/Fe] for C, N, O, Na, Mg, Ca, and Ba from \ddpayne{} & dex \\
e\_[X/Fe]          & Formal uncertainty of each chemical abundance in [X/Fe] & dex \\
$\chi^2_\mathrm{red,A}$ & Reduced-$\chi^2$ of the \ddpaynea{} spectral fitting & \\
$\chi^2_\mathrm{red,G}$ & Reduced-$\chi^2$ of the \ddpayneg{} spectral fitting & \\
\enddata
\tablecomments{
(1) Abundances are calibrated to the literature scale except for [Ca/Fe], as detailed in Section~\ref{subsec:validation-calibration}. \\
(2) Uncertainties are the formal (statistical) errors returned by the \ddpayne{} spectral fitting and are recalculated to include the abundance calibration described in Section~\ref{subsec:validation-calibration} using uncertainty propagation equations. These uncertainties are however noted to be underestimated. \\
(3) $[\mathrm{C/Fe}]$, $[\mathrm{N/Fe}]$, and $[\mathrm{O/Fe}]$ are reported only for stars with $[\mathrm{Fe/H}]_{\rm uncal} > -1.0$~dex. \\
(4) $[\mathrm{Mg/Fe}]$ is reported only for stars with $[\mathrm{Fe/H}]_{\rm uncal} > -1.5$~dex \\
(5) $[\mathrm{Na/Fe}]$ is reported only for stars observed in MUSE non-AO mode.
% (5) \teff{}, \logg{}, \feh{}, $[\mathrm{Ca/Fe}]$, and $[\mathrm{Ba/Fe}]$ are reported for all stars.
}
% \tablecomments{
% This table lists the columns of the machine-readable oMEGACat \ddpayne{} catalog. Abundances are calibrated to literature except for [Ca/Fe], as detailed in Section~\ref{subsec:validation-calibration}. 
% Uncertainties are the formal (statistical) errors returned by the \ddpayne{} spectral fitting with calibrations in Section~\ref{subsec:validation-calibration}. They are noted to be under-estimated. 
% As discussed in Section~\ref{subsec:results-quality}, $[\mathrm{C/Fe}]$, $[\mathrm{N/Fe}]$, and $[\mathrm{O/Fe}]$ are reported only for stars with $[\mathrm{Fe/H}]_{\rm uncal}>-1.0$~dex; $[\mathrm{Mg/Fe}]$ is reported only for stars with $[\mathrm{Fe/H}]_{\rm uncal}>-1.5$~dex; $[\mathrm{Na/Fe}]$ is reported only for stars observed by MUSE non-AO mode; \teff{}, \logg{}, \feh{}, $[\mathrm{Ca/Fe}]$, and $[\mathrm{Ba/Fe}]$ are reported for all stars.}
\label{tab:omegacat_ddpayne_columns}
\end{deluxetable}

% Abundances cut, and remove AO stars when using [Na/Fe]
Based on the reliability assessment of the \ddpayne{} measurements in Section~\ref{subsec:validation-ddp}, we use [C/Fe], [N/Fe], and [O/Fe] only for stars with $\feh_\mathrm{uncal}>-1.0$~dex and $m_\mathrm{F625W}>14$~mag (505 stars), where the magnitude cut-off is applied to remove the effect of CNO variations above the RGB bump in the subsequent analysis \citep[e.g.,][]{Lardo2012A&A}. For [Mg/Fe], we only use stars with $\feh_\mathrm{uncal}>-1.5$~dex (3,563 stars).
For \teff{}, \logg{}, \feh{}, [Na/Fe], [Ca/Fe], and [Ba/Fe], we include all stars across the full metallicity range (7,302 stars).
Additionally, because [Na/Fe] is primarily constrained by the \nad{} region \citepalias{Xiang2019ApJS}, we use only stars observed in non-adaptive-optics (non-AO) mode when analyzing [Na/Fe] (5,881 stars).
In AO-mode observations, the \nad{} is masked to block laser contamination, leading to unreliable [Na/Fe] measurements.
We publish our final oMEGACat \ddpayne{} catalog as an online machine-readable table, including \snrmuse{}, \teff{}, \logg{}, [Fe/H], individual abundance measurements, and literature star identifiers. The column descriptions are provided in Table~\ref{tab:omegacat_ddpayne_columns}.
The table also includes the formal uncertainties of the individual stellar measurements (recalculated due to the abundance calibration described in Section~\ref{subsec:validation-calibration}) for reference, although we note the deviations ($\sigma$) relative to the literature measurements shown in Fig.~\ref{fig:abundances_literature_metallicity_selection} are more representative of the error bars on our individual measurements.

% \textcolor{red}{In the following analysis, for elements that are not covered by \ddpayne{} across the full metallicity range, we incorporate literature data to extend the metallicity coverage.
% Specifically, for [C/Fe], [N/Fe], and [O/Fe] at $\feh_\mathrm{uncal}<-1.0$~dex, and for [Mg/Fe] at $\feh_\mathrm{uncal}<-1.5$~dex, we adopt abundance measurements from \cite{Schiavon2024MNRAS}.}

\subsection{Chemical Abundance Patterns with Metallicity}
\label{subsec:results-abundances-feh}

% Figure. [Fe/H] - [X/Fe] comparison
\begin{figure*}[!ht]
    \centering
    \includegraphics[width=1\linewidth]{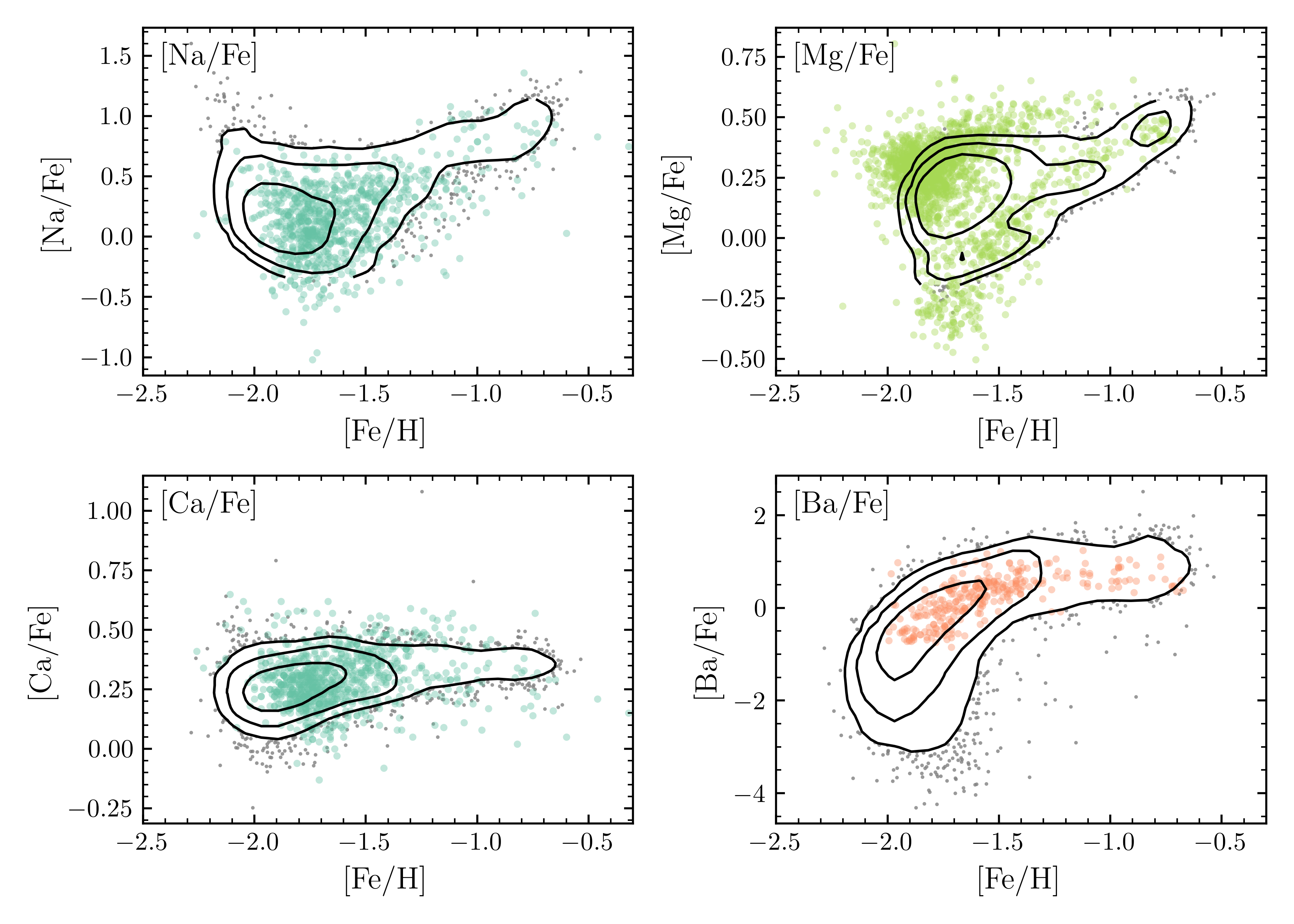}
    \caption{
    Chemical abundances of \omc{} stars as a function of \feh{}.
    The \ddpayne{} measurements are shown as black density contours (50\%, 80\%, and 95\% enclosed probability), while only the lowest-density of stars are over-plotted in gray to show the outer envelope.
    Colored points show literature measurements used as calibration references in Section~\ref{subsec:validation-calibration}, using the same color scheme as Fig.~\ref{fig:abundances_1_1_literature}.
    For the [Mg/Fe]-[Fe/H] relation, only stars with \ddpayne{} $\feh{}_{\rm uncal} > -1.5$~dex are shown.
    \label{fig:xfe_feh}
    }
\end{figure*}

% Introduce [X/Fe]-[Fe/H] figure
Fig.~\ref{fig:xfe_feh} shows the \ddpayne{} chemical abundances as a function of \feh{} (contours and gray points) for [Na/Fe], [Mg/Fe], [Ca/Fe], and [Ba/Fe], with literature measurements shown as colored points.
We focus on these four elements because Section~\ref{subsec:validation-ddp} demonstrated that their MUSE spectral features can be correctly identified by \ddpayne{} in a relatively wide metallicity range.
For literature data, we show only the studies adopted for the metallicity-dependent calibration in Section~\ref{subsec:validation-calibration} with the same color scheme as Fig.~\ref{fig:abundances_1_1_literature}.

% Describe [X/Fe]-[Fe/H] figure
As shown in Fig.~\ref{fig:xfe_feh}, the calibrated \ddpayne{} abundance-metallicity relations for [Na/Fe], [Mg/Fe], and [Ba/Fe] agree well with the literature trends, and the original \ddpayne{} measurements of [Ca/Fe] are also consistent with the literature.
[Na/Fe] increases with metallicity, and its dispersion is consistent with the high-resolution measurements of \cite{Johnson2010ApJ}.
% \textcolor{red}{[Na/Fe] slightly decreases at the lowest metallicities and then increases toward higher \feh{}, with a dispersion broadly consistent with the high-resolution measurements of \cite{Johnson2010ApJ}.
% Compared to the literature data, \ddpayne{} shows a few stars with very high [Na/Fe] values of $\sim1.0$~dex at $\feh{}\sim-2.2$~dex.
% These outliers may be caused by the metallicity-dependent calibration, for which the polynomial correction becomes large at the lowest metallicities.}
[Mg/Fe] exhibits more complex structure in the \ddpayne{} data, while the literature measurements \citep{Schiavon2024MNRAS} show two distinct sequences; both datasets display an overall increasing trend with \feh{}.
The two sequences in literature correspond to the primordial-Mg and Mg-depleted stars associated with the Mg-Al cycle, where Mg is converted into Al only in high-temperature environments ($T \simeq 10^7$~K; \citealp[e.g.,][]{AlvarezGaray2024AA, Mason2026MNRAS}).
Compared to the literature, the \ddpayne{} results partially reveal the two sequences, although they appear to merge at intermediate metallicity ($\feh{}\sim-1.2$~dex). 
This could be due to the lack of stars with $\feh{}_{\rm uncal} < -1.5$~dex in this panel, which affects the contour determination. It also likely reflects the larger [Mg/Fe] uncertainties in the \ddpayne{} estimates and the influence of Milky Way field stars that dominate the \ddpayne{} training set.
The $s$-process element [Ba/Fe] increases continuously with metallicity, reaching super-solar values near $\feh{}=-1.5$~dex, which is also observed in \cite{Marino2011ApJ}.
Compared to the \cite{Marino2011ApJ} data, \ddpayne{} yields [Ba/Fe] values as low as $-2.0$~dex for the most metal-poor stars.
In conclusion, after applying the metallicity-dependent calibration, the \ddpayne{} abundances reproduce the overall abundance-metallicity trends reported in previous literature studies.
% In conclusion, our \ddpayne{} measurements are generally consistent with the abundance-metallicity distributions reported in previous literature studies.

% % Figure. [X/Fe] - [X/Fe] comparison
% \begin{figure}
%     \centering
%     \includegraphics[width=1\linewidth]{xfe_xfe_literature.png}
%     \caption{
%     \textcolor{red}{remove this figure}
%     [Na/Fe]-[O/Fe] relation for stars in the oMEGACat sample (gray) compared to \cite{Johnson2010ApJ} (cyan).
%     The solid line shows the median [Na/Fe] trend in bins of [O/Fe], with shaded regions representing the 16th-84th percentiles.
%     }
%     \label{fig:nafe_ofe}
% \end{figure}

% % Na-O relation
% Although \ddpayne{} can not correctly identify spectral features for [O/Fe] at $\feh<-1.5$~dex, we still examine the [Na/Fe]-[O/Fe] relation as shown in Fig.~\ref{fig:nafe_ofe} because the Na-O anticorrelation is a key diagnostic of multiple stellar populations in globular clusters \citep[e.g.,][]{Gratton2012A&ARv}.
% The \ddpayne{} measurements (gray points and solid line) clearly show the characteristic Na-O anti-correlation, where Na-enhanced stars are O-depleted and vice versa.
% This is expected due to proton-capture nucleosynthesis.
% The agreement with \cite{Johnson2010ApJ} in cyan confirms the reliability of \ddpayne{} abundance measurements.
% Furthermore, this figure includes 5,454 oMEGACat stars, which is an order of magnitude larger than the high-resolution fiber-fed sample of \cite{Johnson2010ApJ}.
% Therefore, the our oMEGACat \ddpayne{} datasets allowing a more comprehensive mapping of the Na-O trend in \omc{}.

\subsection{Chromosome Diagram with Chemical Abundances}
\label{subsec:results-chromosome}

\begin{figure*}[!ht]
    \centering
    \includegraphics[width=0.98\linewidth]{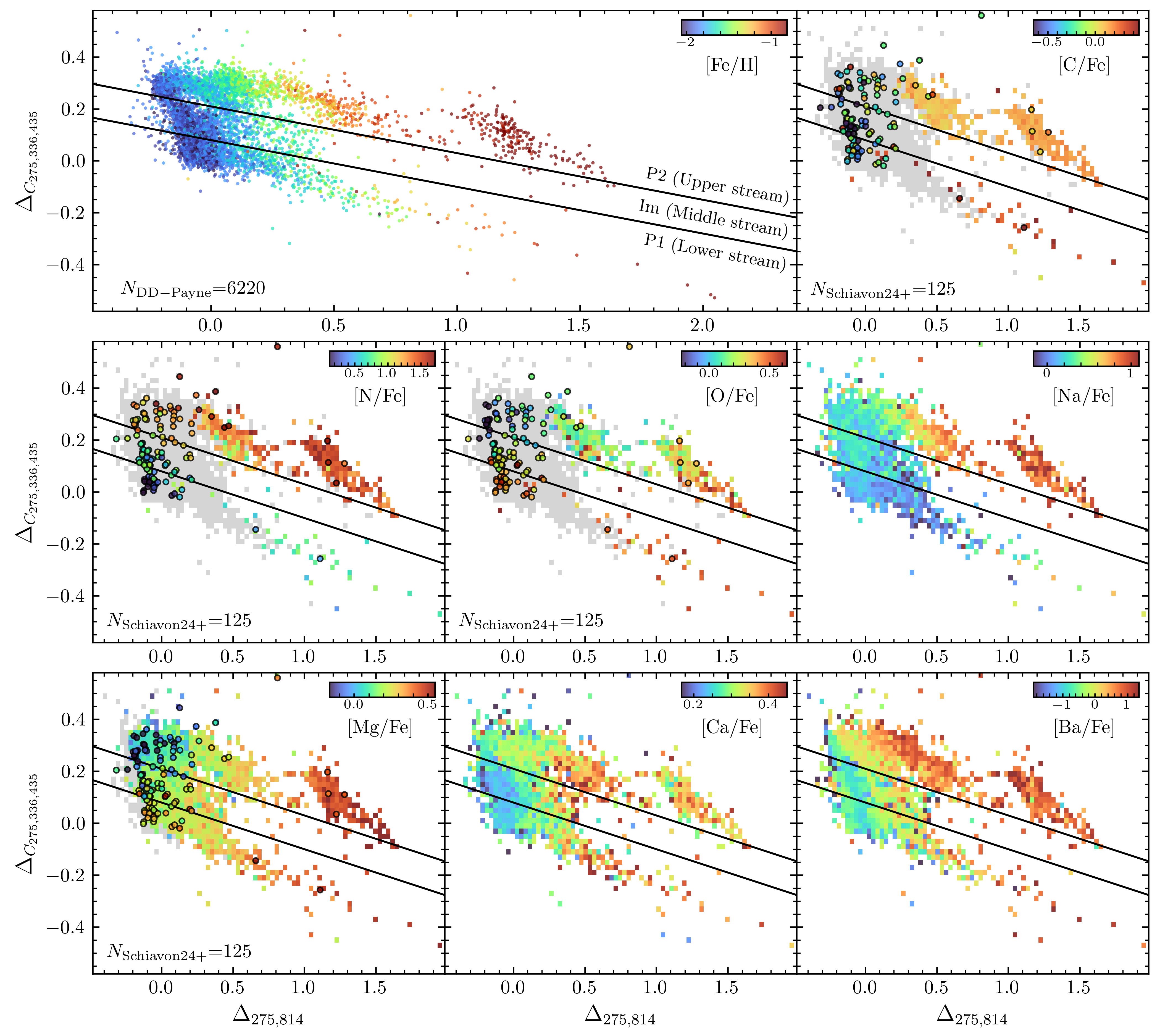}
    \caption{
    Chromosome diagram of \omc{} stars.
    The top-left panel shows individual stars from the oMEGACat \ddpayne{} sample, color-coded by \feh{}.
    The remaining panels show the median \ddpayne{} chemical abundances calculated using stars within each grid cell (0.035 on the x-axis and 0.02 on the y-axis).
    % The oMEGACat \ddpayne{} sample is selected following the criteria described in Section~\ref{subsec:results-quality}.
    The gridded colors are based only on \ddpayne{} measurements.
    Particularly, \ddpayne{} measurements of [C/Fe], [N/Fe], and [O/Fe] are shown only for stars with $\feh{}_{\rm uncal}>-1$~dex, and [Mg/Fe] only for stars with $\feh{}_{\rm uncal}>-1.5$~dex.
    Gray regions in the [C/Fe], [N/Fe], [O/Fe], and [Mg/Fe] panels are areas excluded by these cuts.
    We also plot individual abundance measurements for C, N, O, and Mg from APOGEE DR17 \citep{Schiavon2024MNRAS} to show their variance in the metal-poor regime.
    Black lines indicate the boundaries of the upper (P2), middle (Im), and lower (P1) stellar streams defined in \cite{Clontz2025ApJ}.
    }
    \label{fig:ChM_xfe}
\end{figure*}

% Show chromosome diagram
Our oMEGACat spectra are located in the central regions of the cluster, and thus can be cross-correlated with multi-band HST photometry \citepalias{Haberle2024ApJ} in much greater numbers than any previous high-resolution abundance measurements.  We focus in this section on the chromosome (or pseudo color-color) diagram, which is a powerful tool for analyzing multiple stellar populations with their abundance variations in globular clusters.
Its two axes are constructed from specific photometric bands to trace abundance variations \citep[e.g.,][]{Sbordone2011A&A, Milone2017MNRASb}.
By combining the multi-band \textit{HST} photometry from oMEGACat \citepalias{Haberle2024ApJ} with our \ddpayne{} abundance measurements, we study the distribution of chemical abundances across the chromosome diagram.
In Fig.~\ref{fig:ChM_xfe}, each panel shows the gridded chromosome diagram color-coded by a \ddpayne{}-inferred abundance.
The chromosome diagrams are plotted using the pseudo-colors ($x$ and $y$ coordinates) provided by \cite{Clontz2025ApJ}.
We plot [Fe/H], [Na/Fe], [Ca/Fe], and [Ba/Fe] across the full metallicity range.
[C/Fe], [N/Fe], and [O/Fe] are restricted to $\feh{}_{\rm uncal}>-1.0$~dex and [Mg/Fe] is restricted to $\feh{}_{\rm uncal}>-1.5$~dex.
We note that these metallicity restrictions, together with the non-AO requirement for [Na/Fe] described in Section~\ref{subsec:results-quality}, should be kept in mind when interpreting the C, N, O, Mg, and Na patterns in this figure.
Except for [Fe/H] and [Ca/Fe], which show relatively small offsets from the literature, the other abundances require more significant calibration to the literature scale.
We also plot individual abundance measurements for C, N, O, and Mg from APOGEE DR17 \citep{Schiavon2024MNRAS} as data points to show their values in the metal-poor regime.
The black lines indicate the upper (P2), middle (Im), and lower (P1) stellar stream selections as defined by \cite{Clontz2025ApJ}.

% Findings in chromosome diagram
As shown in Fig.~\ref{fig:ChM_xfe}, [Fe/H] shows a diagonal gradient in the chromosome diagram, and it has a correlation with the $\Delta_{275,814}$ axis.
This is consistent with results from \cite{Nitschai2024ApJ, Latour2025A&A}.
Among the light elements, [C/Fe], [N/Fe], and [O/Fe] from \ddpayne{} display clear abundance variations between the P1 and P2 streams in the metal-rich regime ($\feh{}_{\rm uncal}>-1.0$~dex), although the P1 stream contains only a handful of stars at these metallicities.
In the metal-poor regime ($\feh{}_{\rm uncal}<-1.0$~dex), [N/Fe] and [O/Fe] measurements from APOGEE DR17 (data points) show smooth variations, while the behavior of [C/Fe] is unclear.
[Na/Fe] and [Mg/Fe] from \ddpayne{} exhibit both diagonal and vertical gradients, reflecting the abundance differences in different populations.
The literature [Mg/Fe] measurements are generally consistent with the \ddpayne{} results, with the largest deviation appearing in the upper-left region of the chromosome diagram ($\Delta_{275,814}\sim0$, $\Delta_{C_{275,336,435}}\sim0.3$), corresponding to the metal-poor regime in the P2 stream.
In this region, the literature data points are much bluer than the \ddpayne{} grid, indicating lower [Mg/Fe] values.
Similar to \feh{}, [Ca/Fe] from \ddpayne{} shows a general increasing trend along the x-axis.
Notably, [Ba/Fe] from \ddpayne{} shows a transition near $\Delta_{275,814}\sim0$, corresponding to the lowest-metallicity regime. 
This behavior primarily reflects the rapid increase of [Ba/Fe] with metallicity between $\feh{}\sim -2.0$ and $-1.5$~dex (see Fig.~\ref{fig:xfe_feh}), after which the trend flattens.
We also investigated the uncalibrated [Ba/Fe] from \ddpayne{} and found the same qualitative increase with metallicity, indicating that the trend is not driven solely by the metallicity-dependent calibration.

Compared with traditional high-resolution spectroscopic studies that typically target hundreds of RGB stars, the oMEGACat \ddpayne{} dataset enables statistical analyses of abundance variations across the chromosome diagram for Fe, Na, Mg, Ca, and Ba, where the sample size is much larger.
For C, N, and O, the measurements in the metal-rich regime provide a valuable complement to previous literature studies.
% Compared with traditional high-resolution spectroscopic studies that target only hundreds of RGB stars, the oMEGACat \ddpayne{} dataset allows a far more comprehensive analysis of abundance variations across the chromosome diagram.
The trends observed here are consistent with those found in high-resolution studies of individual stars \citep{Johnson2010ApJ, Marino2011ApJ, Marino2012ApJ, Schiavon2024MNRAS},
% (see Appendix~\ref{appsec:ChM-abundance-lit})
demonstrating that machine-learning-based abundance measurements from MUSE spectra can effectively reproduce key chemical features of \omc{}'s multiple stellar populations.

\section{Discussion} 
\label{sec:disscuss}

\subsection{Abundance Variations in P1, Im, and P2 Streams}
\label{subsec:discuss-stream}

\begin{figure*}[!ht]
    \centering
    \includegraphics[width=1\linewidth]{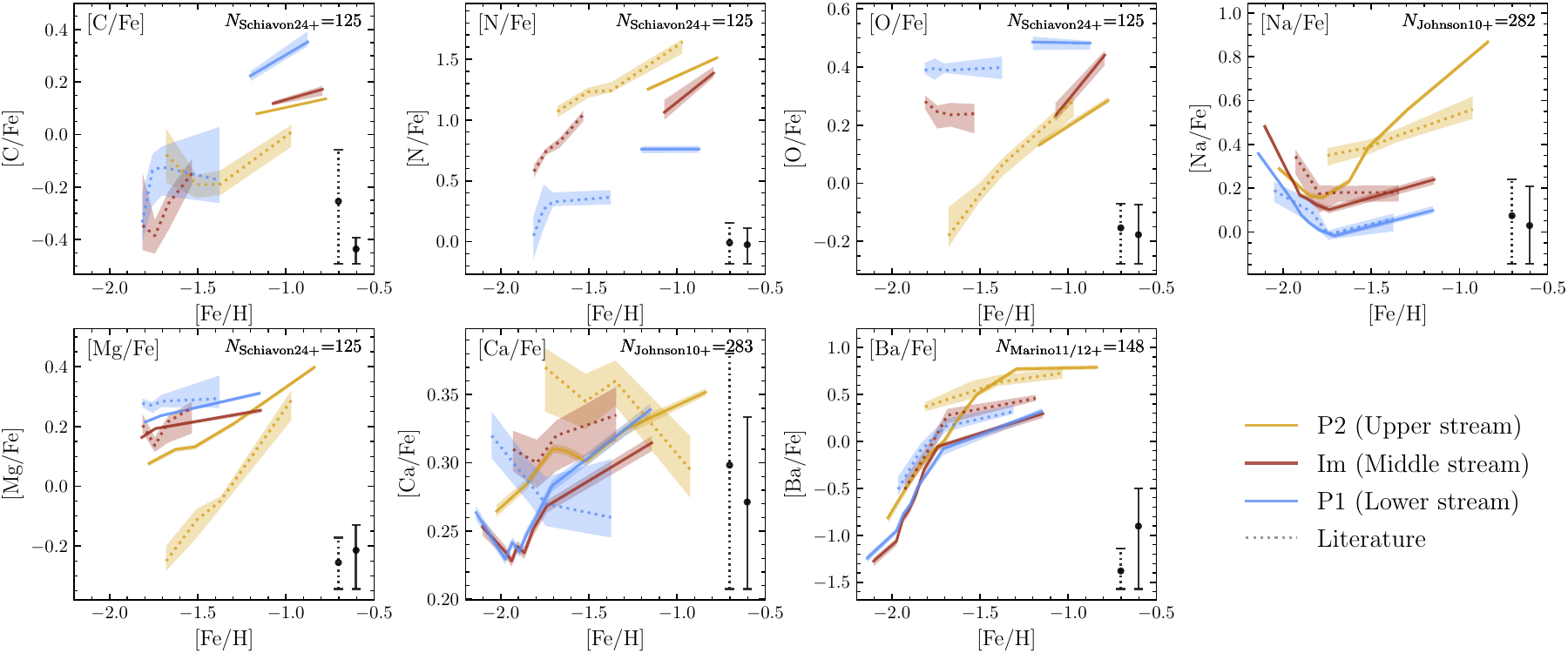}
    \caption{
    Chemical abundances as a function of \feh{}.
    Abundances for the three streams are noted as Upper (P2), Middle (Im), and Lower (P1), as defined by the black solid lines in Fig.~\ref{fig:ChM_xfe}. 
    We plot the median trend for each stream, with shaded regions representing the standard error of the mean (SEM). 
    All \feh{} bins are constructed separately for each stream and dataset to contain equal numbers of stars; the number per bin ranges from 7 to 332 depending on the element and sample size.
    Solid lines represent \ddpayne{} measurements (selected via criteria in Section~\ref{subsec:results-quality}).
    Dotted lines indicate literature results with the reference and number of stars written in the top right corner of each panel.
    The reference error bars show the median 16th--84th percentile abundance range across the individual bins, with solid and dotted error bars corresponding to the \ddpayne{} and literature samples, respectively.
    }
    \label{fig:xfe_feh_stream}
\end{figure*}

As shown in Fig.~\ref{fig:ChM_xfe}, the distribution of stars on the chromosome diagram clearly reveals the presence of multiple stellar populations in \omc{}. 
The most common classification divides stars into two primary groups, P1 and P2, which has been widely adopted to study abundance variations in globular clusters \citep[e.g.,][]{Marino2011ApJ, Marino2012ApJ}. 
Recent studies identified another ``Intermediate'' (Im) population having a narrower metallicity spread than P1 and P2 \citep{Marino2019MNRAS,Clontz2025ApJ, Mason2026MNRAS, Dondoglio2026A&A} although the exact naming of these three different populations varies between these studies.
\citet{Clontz2025ApJ} compared the chromosome diagram with theoretical isochrones and demonstrated that the P2 population is helium-enhanced across all metallicities compared to P1 and Im  ($\Delta Y_{\rm min} \ge 0.11$).
In this subsection, we adopt the classification of three primary populations from \citet{Clontz2025ApJ}, dividing our oMEGACat \ddpayne{} sample into Upper (P2), Middle (Intermediate), and Lower (P1) streams to investigate their chemical abundance distinctions.

The results are shown in Fig.~\ref{fig:xfe_feh_stream}, where we compare the median abundance trends with error bars calculated using the standard error of the mean (SEM; shaded regions) of the three streams using \ddpayne{} measurements (solid lines) based on the selection criteria in Section~\ref{subsec:results-quality}, together with literature data (dotted lines).
For consistency, the literature datasets used for each element are the same as those adopted for the metallicity-dependent calibration in Section~\ref{subsec:validation-calibration}.
We note that the SEMs quantify the uncertainty on the median abundance trends rather than the abundance dispersion within each stream.
To illustrate the latter, we additionally show reference error bars in each panel representing the median 16th--84th percentile abundance range across the individual bins, separately for the \ddpayne{} and literature samples.
These ranges are substantially larger than the SEMs and provide a more direct representation of the observed abundance spread within each stream.
The boundaries defining the three streams are indicated by the black lines in Fig.~\ref{fig:ChM_xfe}.

For C, N, and O, as shown in Fig.~\ref{fig:xfe_feh_stream}, our \ddpayne{} measurements extend the literature abundance-metallicity relations to higher metallicities.
From the combined datasets, [N/Fe] shows a clear increase with metallicity for P2 and Im, whereas P1 is relatively flat with a mild overall increase.
For [O/Fe], P2 shows the strongest increase with metallicity, Im exhibits a weaker increasing trend, and P1 is mostly constant.
Both [N/Fe] and [O/Fe] show clear offsets between the three streams. 
[N/Fe] increases from P1, Im to P2, whereas [O/Fe] follows the opposite ordering, with P2 being the most O-depleted and P1 the most O-enhanced. This ordering illustrates the N-O anti-correlation across all metallicities.
Consistent with \citet{Marino2012ApJ}, we do not find a strong internal [C/Fe]-metallicity trend within any individual stream. 
The literature [C/Fe] measurements are highly scattered with large SEM, which is also seen in the chromosome diagram in Fig.~\ref{fig:ChM_xfe}.
The newly included \ddpayne{} measurements suggest that the overall [C/Fe] may increase toward the metal-rich end, but the significance is low because the \ddpayne{} P1 and Im samples in this regime contain only 39 and 16 stars, respectively. Future investigations with more accurate [C/Fe] measurements are needed to confirm this behavior.

Furthermore, [Na/Fe] also shows distinct trends among the three streams: it decreases at the lowest metallicities and then increases toward higher \feh{}, with P2 being more Na-enriched than P1 and Im.
The \ddpayne{} trend for P2 also appears to increase more steeply than in the literature.
[Mg/Fe] is mildly increasing with metallicity in P1 and Im, while P2 is Mg-depleted but shows a steeper increase toward higher metallicities.
Although P1 and Im show consistent trends between \ddpayne{} and the literature, the P2 stream in the literature extends to lower values ([Mg/Fe]~$\sim-0.2$~dex) than in the \ddpayne{} measurements. 
This discrepancy is likely caused by the \ddpayne{} training set, which is dominated by Milky Way field stars and therefore affects the abundance estimates for the Mg-depleted stars.
Both the \ddpayne{} and literature measurements of [Ca/Fe] lie within a similarly narrow range (0.23-0.39~dex), although the P2 population in the literature shows a decreasing trend with metallicity.

For the $s$-process element [Ba/Fe], all three streams in Fig.~\ref{fig:xfe_feh_stream} show an increasing trend with \feh{} which flattens toward the metal-rich end.
Notably, despite small offsets between the \ddpayne{} (solid lines) and \citet{Marino2011ApJ} (dotted lines) measurements, both datasets show that P2 reaches a higher [Ba/Fe] value than P1 and Im.
This feature was not detected by \citet{Marino2012ApJ}, who found no significant difference in [La/Fe]--another heavy $s$-process element--between their P1 and P2 populations.
However, using [La/Fe] from both \citet{Marino2012ApJ} and \citet{Schiavon2024MNRAS}, we find that P2 is indeed more enhanced than P1 and Im under our population definition, at a level comparable to [Ba/Fe].
Therefore, we believe that the enhanced $s$-process enrichment in P2 stars is a real signature, and the non-detection of \citet{Marino2012ApJ} is due to the different definitions of P1 and P2, as their classification is based on the Na-O anti-correlation.
Ba enhancement in P2-like stars has also been reported in other globular clusters \citep[e.g.,][]{Kacharov2013A&A, Marino2019MNRAS}.
Our finding is further supported by recent studies of \omc{} \citep{Mason2026MNRAS, Dondoglio2026A&A}, who reported similar results using $s$-process elements [Ce/Fe], [Ba/Fe], and [La/Fe].
% However, we also use the [Ba/Fe] measurements from \citet{Marino2011ApJ} (dotted line in Fig.~\ref{fig:xfe_feh_stream}) and [La/Fe] from both \cite{Marino2012ApJ} and \cite{Schiavon2024MNRAS}, and find P2 is more enhanced than P1 and Im. 
Given that $s$-process elements are primarily produced by low-mass AGB stars ($1.5$-$3~M_{\odot}$) over timescales of $0.3$-$2$~Gyr \citep{Smith2000AJ}, which is comparable to the $\sim$1~Gyr age difference between P1 and P2 \citep{Clontz2024ApJ}, the enhanced [Ba/Fe] in P2 could be naturally explained if P2 formed from gas enriched by the AGB ejecta of the earlier P1 stars.

\subsection{Abundance Variations in 14 Subpopulations}
\label{subsec:discuss-subpopulations}

\begin{figure*}[!ht]
    \centering
    \includegraphics[width=1\linewidth]{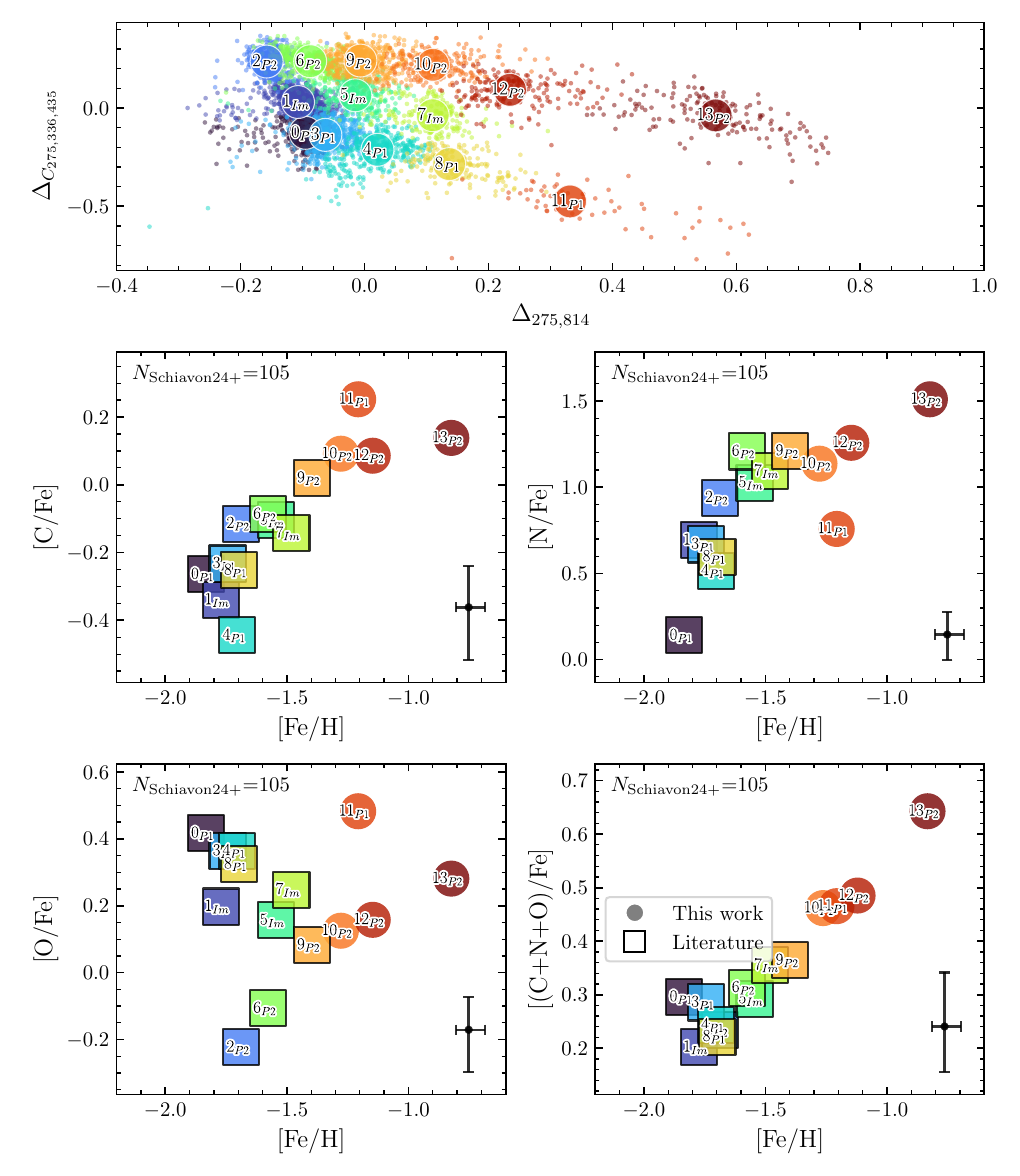}
    \caption{
    Chemical abundances of C, N, O, and (C+N+O) as a function of \feh{} for subpopulations defined by \cite{Clontz2026ApJ}.
    The top panel shows the chromosome diagram based on the photometry from \cite{Clontz2026ApJ}.
    For the remaining panels, we plot the median abundance as a function of metallicity for each element to illustrate the abundance-metallicity relations.
    Circles represent the \ddpayne{} measurements following the selection criteria in Section~\ref{subsec:results-quality}, and squares are literature results, with the number of literature stars used written in the top left corner.
    The black reference error bar shows the median 16th--84th percentile range in [Fe/H] and abundance across the individual subpopulations plotted in each panel; these are not uncertainties on the median data points, but instead illustrate the typical observed abundance and metallicity spread within the subpopulations.
    }
    \label{fig:xfe_feh_subpops_CNO}
\end{figure*}

\begin{figure*}[!ht]
    \centering
    \includegraphics[width=1\linewidth]{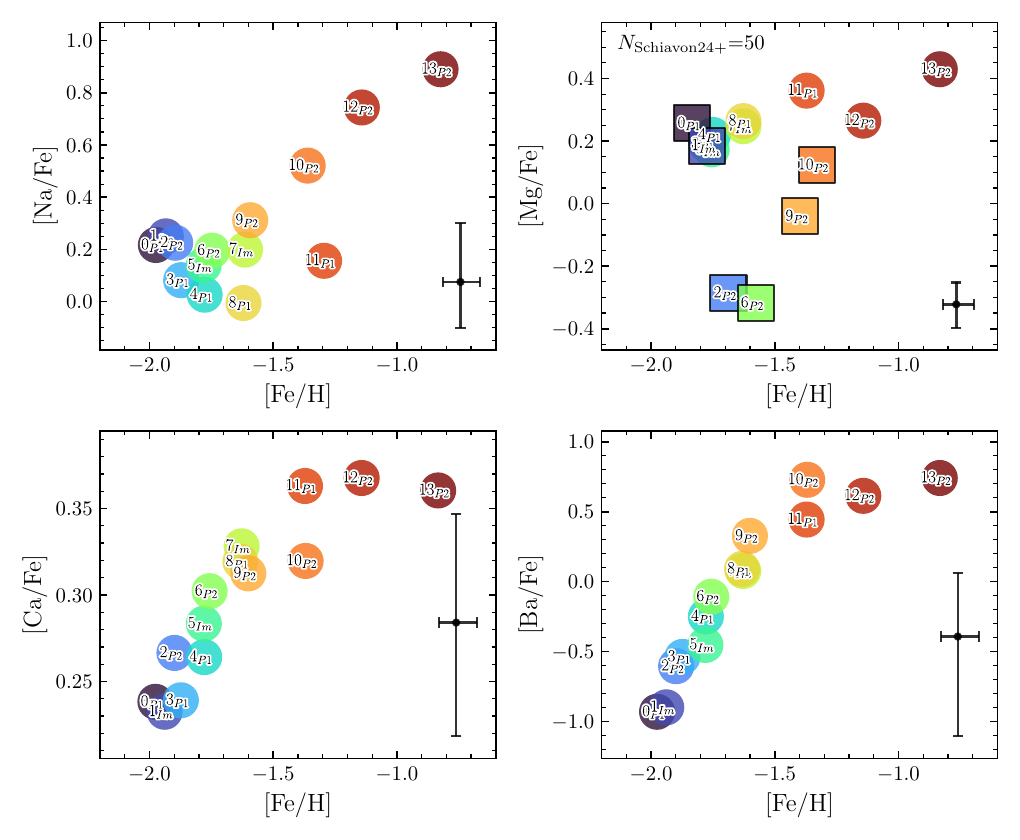}
    \caption{
    Same as Fig.~\ref{fig:xfe_feh_subpops_CNO} but for Na, Mg, Ca, and Ba.
    }
    \label{fig:xfe_feh_subpops_NaMgBa}
\end{figure*}

\begin{figure}
    \centering
    \includegraphics[width=1.0\linewidth]{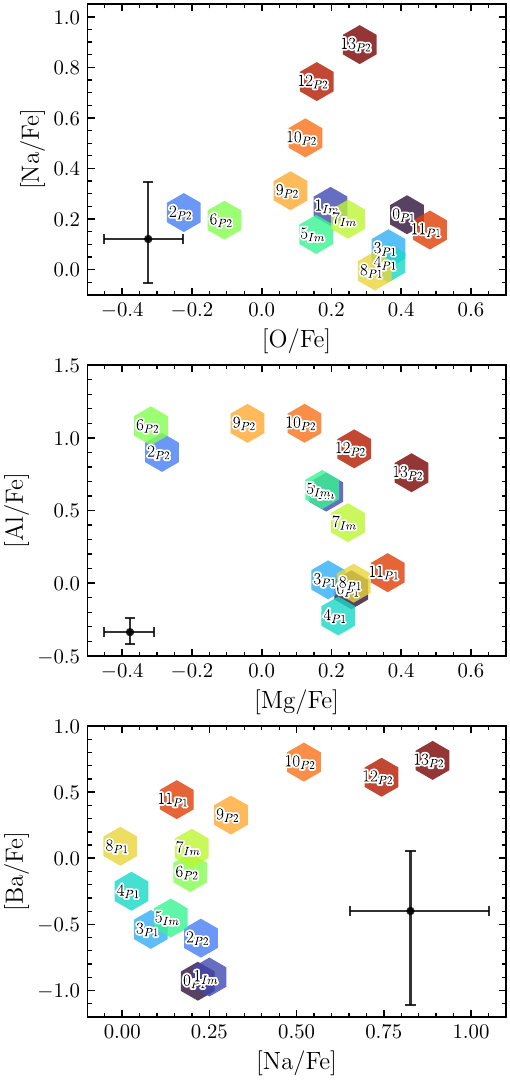}
    \caption{
    Na--O, Mg--Al, and Na--Ba abundance relations for the 14 subpopulations identified by \citet{Clontz2026ApJ}.
    [O/Fe] and [Mg/Fe] are derived from both the
    \ddpayne{} measurements and \citet{Schiavon2024MNRAS}, while [Al/Fe]
    is taken solely from \citet{Schiavon2024MNRAS}.
    [Na/Fe] and [Ba/Fe] are taken entirely from the oMEGACat \ddpayne{} measurements.
    All corresponding abundance values are listed in
    Table~\ref{tab:subpop_abundances}.
    Unlike Fig.~\ref{fig:xfe_feh_subpops_NaMgBa}, the abundances shown on the two axes of each panel may be derived from different data sources.
    We therefore use a hexagon symbol for each subpopulation to denote these combined measurements.
    The black reference error bars are plotted in the same
    way as in Fig.~\ref{fig:xfe_feh_subpops_CNO}.
    }
    \label{fig:xfe_xfe_subpops}
\end{figure}

% \begin{deluxetable*}{rccccccccc}
% \tablecaption{Median chemical abundances of \omc{} subpopulations\label{tab:subpop_abundances}}
% \tablehead{
% \colhead{Pop.} &
% \colhead{ChD.} &
% \colhead{[Fe/H]} &
% \colhead{[C/Fe]} &
% \colhead{[N/Fe]} &
% \colhead{[O/Fe]} &
% \colhead{[C+N+O/Fe]} &
% \colhead{[Na/Fe]} &
% \colhead{[Mg/Fe]} &
% \colhead{[Ba/Fe]}
% }
% \startdata
% 0  & P1 & -1.976 & -0.265 & 0.141 & 0.417 & 0.296 & 0.159 & 0.257 & -0.930 \\
% 1  & Im & -1.939 & -0.340 & 0.692 & 0.197 & 0.202 & 0.189 & 0.185 & -0.899 \\
% 2  & P2 & -1.899 & -0.115 & 0.940 & -0.223 & 0.234 & 0.184 & -0.286 & -0.605 \\
% 3  & P1 & -1.873 & -0.232 & 0.668 & 0.364 & 0.286 & 0.035 & 0.192 & -0.538 \\
% 4  & P1 & -1.778 & -0.444 & 0.515 & 0.364 & 0.243 & -0.001 & 0.220 & -0.249 \\
% 5  & Im & -1.780 & -0.105 & 1.026 & 0.156 & 0.292 & 0.130 & 0.175 & -0.453 \\
% 6  & P2 & -1.757 & -0.087 & 1.208 & -0.107 & 0.313 & 0.185 & -0.318 & -0.108 \\
% 7  & Im & -1.627 & -0.138 & 1.091 & 0.250 & 0.360 & 0.211 & 0.248 & 0.076 \\
% 8  & P1 & -1.632 & -0.253 & 0.596 & 0.324 & 0.222 & 0.010 & 0.264 & 0.091 \\
% 9  & P2 & -1.600 & 0.020  & 1.211 & 0.083 & 0.364 & 0.331 & -0.041 & 0.328 \\
% 10 & P2 & -1.369 & 0.082  & 1.303 & 0.096 & 0.454 & 0.533 & 0.123 & 0.728 \\
% 11 & P1 & -1.371 & 0.228  & 0.794 & 0.478 & 0.449 & 0.158 & 0.363 & 0.444 \\
% 12 & P2 & -1.142 & 0.075  & 1.328 & 0.162 & 0.488 & 0.786 & 0.268 & 0.614 \\
% 13 & P2 & -0.833 & 0.126  & 1.505 & 0.293 & 0.624 & 0.962 & 0.437 & 0.741 \\
% \enddata
% \end{deluxetable*}

\begin{deluxetable*}{cccccccccccccc}[!ht]
\tablecaption{Number of stars and median chemical abundances of \omc{} subpopulations \label{tab:subpop_abundances}}
\tablehead{
\colhead{Pop.} &
\colhead{ChD.} &
\colhead{$N_{\rm DD\text{-}Payne}$} &
\colhead{$N_{\rm S24+}$} &
\colhead{[Fe/H]} &
\colhead{[C/Fe]} &
\colhead{[N/Fe]} &
\colhead{[O/Fe]} &
\colhead{[(C+N+O)/Fe]} &
\colhead{[Na/Fe]} &
\colhead{[Mg/Fe]} &
\colhead{[Al/Fe]} &
\colhead{[Ca/Fe]} &
\colhead{[Ba/Fe]} \\
\colhead{(1)} &
\colhead{(2)} &
\colhead{(3)} &
\colhead{(4)} &
\colhead{(5)} &
\colhead{(6)} &
\colhead{(7)} &
\colhead{(8)} &
\colhead{(9)} &
\colhead{(10)} &
\colhead{(11)} &
\colhead{(12)} &
\colhead{(13)} &
\colhead{(14)}
}
\startdata
0  & P1 & 550 & 14 & -1.98 & -0.26* & 0.14* & 0.42* & 0.30* & 0.22 & 0.26* & -0.05* & 0.24 & -0.93 \\
1  & Im & 724 & 13 & -1.94 & -0.34* & 0.69* & 0.20* & 0.20* & 0.25 & 0.18* & 0.63* & 0.23 & -0.90 \\
2  & P2 & 528 & 7  & -1.90 & -0.11* & 0.94* & -0.22* & 0.23* & 0.23 & -0.29* & 0.90* & 0.27 & -0.60 \\
3  & P1 & 594 & 24 & -1.87 & -0.23* & 0.67* & 0.36* & 0.29* & 0.08 & 0.19  & 0.02* & 0.24 & -0.54 \\
4  & P1 & 378 & 17 & -1.78 & -0.44* & 0.52* & 0.36* & 0.24* & 0.03 & 0.22  & -0.23* & 0.26 & -0.25 \\
5  & Im & 323 & 6  & -1.78 & -0.10* & 1.03* & 0.16* & 0.29* & 0.14 & 0.17  & 0.64* & 0.28 & -0.45 \\
6  & P2 & 494 & 5  & -1.76 & -0.09* & 1.21* & -0.11* & 0.31* & 0.20 & -0.32* & 1.08* & 0.30 & -0.11 \\
7  & Im & 168 & 7  & -1.63 & -0.14* & 1.09* & 0.25* & 0.35* & 0.20 & 0.25  & 0.42* & 0.33 & 0.08 \\
8  & P1 & 177 & 5  & -1.63 & -0.25* & 0.60* & 0.32* & 0.22* & 0.01 & 0.26  & 0.00* & 0.32 & 0.09 \\
9  & P2 & 509 & 7  & -1.60 & 0.02* & 1.21* & 0.08* & 0.36* & 0.31 & -0.04* & 1.10* & 0.31 & 0.33 \\
10 & P2 & 216 & 4  & -1.37 & 0.09  & 1.14  & 0.13  & 0.46  & 0.52 & 0.12* & 1.10* & 0.32 & 0.73 \\
11 & P1 & 62  & 1  & -1.37 & 0.25  & 0.76  & 0.48  & 0.47  & 0.16 & 0.36  & 0.07* & 0.36 & 0.44 \\
12 & P2 & 184 & 2  & -1.14 & 0.09  & 1.26  & 0.16  & 0.49  & 0.74 & 0.27  & 0.92* & 0.37 & 0.61 \\
13 & P2 & 157 & 1  & -0.83 & 0.14  & 1.51  & 0.28  & 0.64  & 0.89 & 0.43  & 0.76* & 0.36 & 0.74 \\
\enddata
\tablecomments{
Column (1): Subpopulation ID. Column (2): Chromosome-diagram class (P1, P2, or Im).
Columns (3)--(4): Number of stars contributing to each subpopulation from \ddpayne{} (this work) and \cite{Schiavon2024MNRAS}, respectively.
Columns (5)--(14): Median abundance of [Fe/H], [C/Fe], [N/Fe], [O/Fe], [C+N+O/Fe], [Na/Fe], [Mg/Fe], [Al/Fe], [Ca/Fe], and [Ba/Fe]. Values are rounded to two decimal places; Population abundances marked with an asterisk (*) are from the literature.}
\end{deluxetable*}

The recent study by \cite{Clontz2026ApJ} employed a clustering algorithm using both photometric chromosome diagram information and spectroscopic metallicity to identify 14 subpopulations in \omc{} from the red giant branch to below the main sequence turnoff using the oMEGACat dataset.
Some results from our abundance measurements were already presented in that work.
In particular, the [Na/Fe] abundances of red giant and subgiant stars show that the subpopulations identified from the chromosome diagram can be consistently traced onto the subgiant branch. This correspondence enables subgiant ages \citep{Clontz2024ApJ} to be inferred for each subpopulation.
We focus on chemical abundance trends with [Fe/H] in this subsection and discuss trends with age in the next subsection.

In this section, we combine the \citet{Clontz2026ApJ} subpopulation classifications with our \ddpayne{} and literature abundance measurements to investigate chemical abundance variations among different subpopulations.
The results are presented in Fig.~\ref{fig:xfe_feh_subpops_CNO} and Fig.~\ref{fig:xfe_feh_subpops_NaMgBa}, the top panel of Fig.~\ref{fig:xfe_feh_subpops_CNO} shows the RGB chromosome diagram labeled by subpopulations from \cite{Clontz2026ApJ}, and the remaining panels in these two figures illustrate the median abundance trends with metallicity.
Here, circles represent oMEGACat \ddpayne{} samples and squares represent literature data.
The literature source used for each element is the same as those in Fig.~\ref{fig:xfe_feh_stream}.
For [Na/Fe], [Ca/Fe], and [Ba/Fe], we use abundance measurements from \ddpayne{} for all subpopulations.
For [C/Fe], [N/Fe], and [O/Fe], we adopt \ddpayne{} measurements for the metal-rich subpopulations (10-13), and use literature measurements for the remaining subpopulations.
For [Mg/Fe], we use literature measurements for subpopulations 0 and 1 due to their low metallicities, where \ddpayne{} is less reliable, as well as for P2 subpopulations 2, 6, 9, and 10, where \ddpayne{} measurements show inconsistencies with the literature (as shown in Fig.~\ref{fig:xfe_feh_stream}).
We list the number of \ddpayne{} and literature stars, along with the median chemical abundances used in Fig.~\ref{fig:xfe_feh_subpops_CNO} and Fig.~\ref{fig:xfe_feh_subpops_NaMgBa}, for all 14 subpopulations in Table~\ref{tab:subpop_abundances}.
We combine \ddpayne{} and literature measurements here to obtain abundance coverage across the full metallicity range of \omc{}.
Although the \ddpayne{} uncertainties are larger than those of high-resolution literature measurements, the \ddpayne{} sample contains many more stars and has been calibrated to the literature abundance scale. 
Moreover, the adopted measurements have been tested to be reliable in Sections~\ref{sec:validation}--\ref{sec:results}.
Therefore, we do not expect this combination to significantly affect the qualitative subpopulation trends.
For this reason, in the following discussion we focus on the combined abundance trends rather than separating the \ddpayne{} and literature results.
To illustrate the typical abundance spread within the subpopulations, we calculate the 16th--84th percentile abundance range for each subpopulation and show the median of these ranges as the black reference error bar in each panel.
%To better illustrate the significance of these results, we show the 16th-84th percentile abundance ranges as error bars for each subpopulation in the following figures.}

As shown in Fig.~\ref{fig:xfe_feh_subpops_CNO} and Fig.~\ref{fig:xfe_feh_subpops_NaMgBa}, [C/Fe], [N/Fe], [Na/Fe], [Ca/Fe], and [Ba/Fe] all show a nearly monotonic increase with metallicity across all subpopulations.
Overall, most P2 subpopulations (2, 6, 9, 10, 12, and 13) show enhanced enrichment in [N/Fe] and [Na/Fe], and generally higher [Ba/Fe], consistent with the three-stream trends in Fig.~\ref{fig:xfe_feh_stream}.
% The expected enhancements of [N/Fe], [Na/Fe], and [Ba/Fe] in the P2 subpopulations are observed and are consistent with the three-stream trends in Fig.~\ref{fig:xfe_feh_stream}. 
Moreover, using the combined \ddpayne{} and literature C, N, and O measurements, we compute the total C+N+O and find a tight increasing trend between [(C+N+O)/Fe] and metallicity, with no significant differences between P1, Im, and P2, consistent with the results of \citet{Marino2012ApJ}. 

The $\alpha$-elements [O/Fe] and [Mg/Fe] based on hybrid literature and \ddpayne{} measurements show more complex patterns in Fig.~\ref{fig:xfe_feh_subpops_CNO} and Fig.~\ref{fig:xfe_feh_subpops_NaMgBa}.
[O/Fe] shows a decreasing trend with metallicity from the most metal-poor to the metal-rich P1 subpopulations (subpops 0, 3, 4, and 8). The Im populations (1, 5, 7) continue this trend toward higher metallicities.  In strong contrast with the P1 populations, but consistent with Fig.~\ref{fig:xfe_feh_stream}, the P2 subpopulations show a steep increase of [O/Fe] with metallicity.
Combining the \ddpayne{} [Na/Fe] measurements with the hybrid [O/Fe] compilation, we find that all the P2 subpopulations are Na-enhanced and O-depleted compared to P1, exhibiting the classic Na-O anti-correlation expected from high-temperature hydrogen-burning \citep[e.g.,][]{Carretta2009A&A}, consistent with previous studies \citep[e.g.][]{Johnson2010ApJ, Marino2011ApJ}. 
We further confirm this by plotting the median [Na/Fe] and [O/Fe] for all subpopulations in Fig.~\ref{fig:xfe_xfe_subpops}.
Similar to [O/Fe], [Mg/Fe] of all the P2 subpopulations are depleted which is consistent with Fig.~\ref{fig:xfe_feh_stream}. However, there is no decreasing trend of [Mg/Fe] with metallicity in P1 and Im subpopulations. Instead, they are located at very similar places with $\feh{}\sim-1.75$~dex and $\mgfe{}\sim0.2$~dex.
To investigate the potential Mg-Al anti-correlation, we take [Al/Fe] measurements from \citet{Schiavon2024MNRAS} and plot the median [Al/Fe] and hybrid [Mg/Fe] distributions for all subpopulations in the middle panel of Fig.~\ref{fig:xfe_xfe_subpops}. The results are also consistent with the literature \citep[e.g.,][]{AlvarezGaray2024AA, Mason2026MNRAS}.

Fig.~\ref{fig:xfe_xfe_subpops} also shows the relation between
the median [Na/Fe] and [Ba/Fe] abundances of the 14 subpopulations,
providing an additional perspective on the connection between
proton-capture and $s$-process enrichment.
Previous studies of individual stars in most Galactic globular clusters found that large variations in Na are generally not accompanied by comparable variations in Ba, suggesting that proton-capture and neutron-capture enrichment are usually decoupled \citep{Smith2008PASP, D'Orazi2013MNRAS}.
However, connections between Na and Ba have been identified in some chemically complex clusters.
In NGC~1851, Ba-rich stars are preferentially Na-rich \citep{Carretta2011A&A, Gratton2012A&A}, while the $s$-process-rich population of M22 is also more Na-rich on average than its $s$-process-poor counterpart \citep{Marino2009A&A}.
In M75, \citet{Kacharov2013A&A} found that the two most Na-rich stars are also significantly Ba-enhanced, indicating a connection between proton-capture and $s$-process enrichment in its chemically distinct populations.
As shown in the bottom panel of Fig.~\ref{fig:xfe_xfe_subpops}, the 14 subpopulations of \omc{} similarly trace an overall increase in both [Na/Fe] and [Ba/Fe], but follow a different sequence from that seen in M75.
In \omc{}, the subpopulations become strongly enhanced in Ba before reaching their highest Na abundances, whereas the M75 populations show a larger Na enhancement before the strongest Ba enrichment.
This difference indicates that the relative enrichment of proton-capture and $s$-process elements proceeded differently in the two clusters, further highlighting the unusually complex chemical-enrichment history of \omc{}.

%Regardless of the continuous enrichment trends and typical anti-correlations described above that normally suggest a common formation environment for the P1 and P2 subpopulations, 

One notable exception to many of the trends discussed above is the most metal-rich P1 subpopulation (11), which is an outlier on the diagrams in several abundances. 
Population 11 follows some trends, such as those of Mg, Ba, and C+N+O with metallicity; it appears more like a discrete group with no clear continuity in C, N, O, and Na. 
Although this population contains fewer \ddpayne{} stars (62) than the other subpopulations with \ddpayne{} measurements, its 16th-84th percentile abundance range is comparable to those of the other subpopulations.
Therefore, its distinct abundance pattern may indicate a different formation pathway for some subpopulations, which we will discuss further in Section~\ref{sec:discuss-senario}.

% \textcolor{red}{The rapid s-process enrichment could be caused by massive rotating stars (cite https://ui.adsabs.harvard.edu/abs/2014ApJ...795...34S/abstract, https://ui.adsabs.harvard.edu/abs/2016MNRAS.456.1803F/abstract). As discussed in https://arxiv.org/abs/2509.16719}

\subsection{Relation between Abundances and Age}
\label{subsec:discuss-xfe-age}

\begin{figure*}[!ht]
    \centering
    \includegraphics[width=1\linewidth]{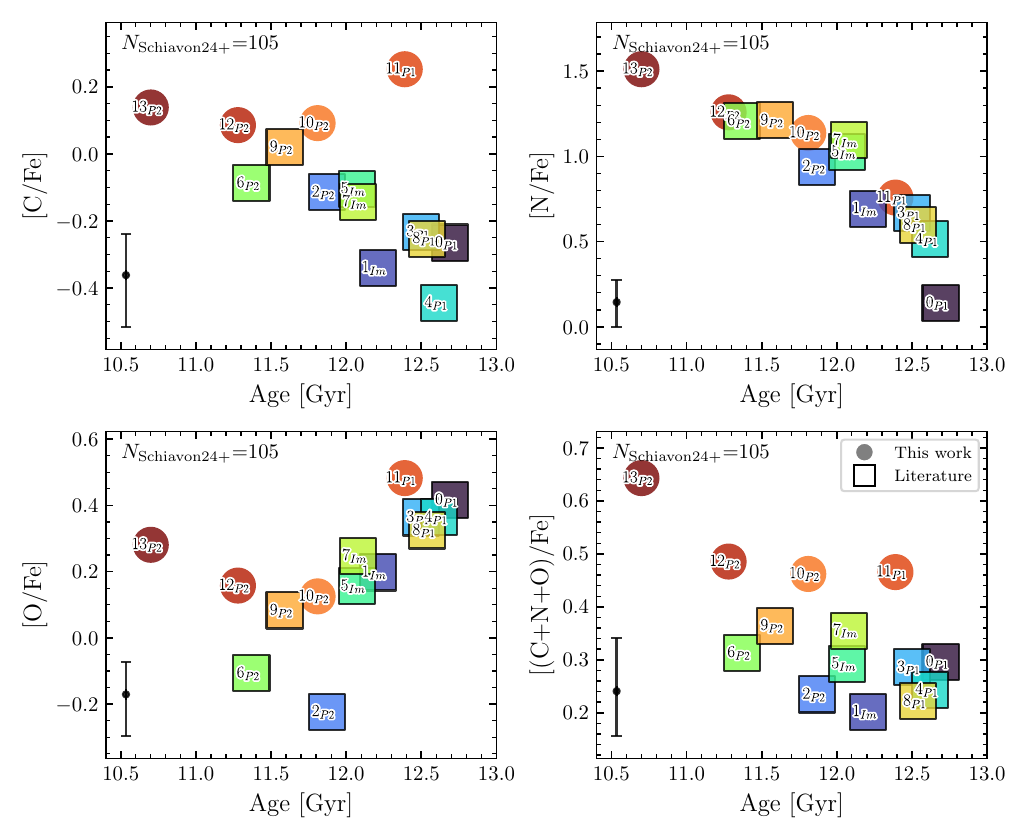}
    \caption{
    Chemical abundances of C, N, O, and (C+N+O) as a function of age for different subpopulations, plotted in the same way as Fig.~\ref{fig:xfe_feh_subpops_CNO} and Fig.~\ref{fig:xfe_feh_subpops_NaMgBa}.
    The median stellar ages 
    for each population are adopted from \cite{Clontz2026ApJ}.
    The black reference error bars are plotted in the same way as in Fig.~\ref{fig:xfe_feh_subpops_CNO}, showing the median abundance range for subpopulations along the y-axis.
    }
    \label{fig:xfe_age_subpops_CNO}
\end{figure*}

% \begin{figure}
%     \centering
%     \includegraphics[width=1.0\linewidth]{c+n+o_age_subpops_with_lit_mask_select.pdf}
%     \caption{
%     [(C+N+O)/Fe] as a function of age for \ddpayne{} and literature results of RGB stars in different subpopulations, plotted in the same way as Fig.~\ref{fig:xfe_age_subpops}.
%     }
%     \label{fig:cno_age}
% \end{figure}

\begin{figure*}
    \centering
    \includegraphics[width=1.0\linewidth]{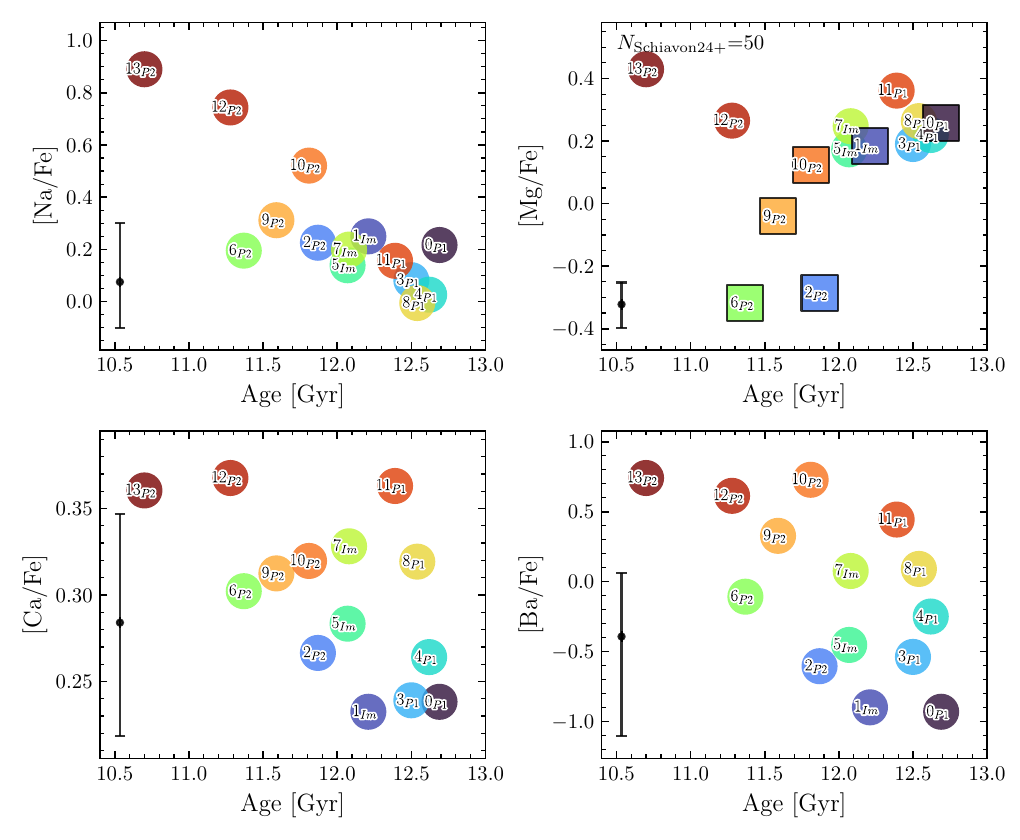}
    \caption{
    Same as Fig.~\ref{fig:xfe_age_subpops_CNO} but for Na, Mg, Ca, and Ba.
    }
    \label{fig:xfe_age_subpops_NaMgBa}
\end{figure*}

In this section, we investigate the abundance-age relation for the subpopulations by combining results from \citet{Clontz2024ApJ} and \citet{Clontz2026ApJ}.  We focus especially on the behavior of P1, Im, and P2 groups given the $\sim$1~Gyr age gap between P1 and P2 reported in \citet{Clontz2026ApJ}.
The results are shown in Fig.~\ref{fig:xfe_age_subpops_CNO} and Fig.~\ref{fig:xfe_age_subpops_NaMgBa}.
For the light elements, we find that the 14 subpopulations based on hybrid literature and \ddpayne{} measurements show the tightest relation between [N/Fe] and age, with no significant differences between P1, Im, and P2. 
Similar to [N/Fe], [C/Fe] increases toward younger ages, except for the most metal-rich P1 subpopulation (11).
% [C/Fe], like [N/Fe], also increases with time past since the formation, except for the most metal-rich P1 subpopulation (11). 
Both [O/Fe] and [Mg/Fe] based on hybrid datasets generally decrease with younger ages except for the two most-metal-rich P2 populations (12 and 13), which are enriched in both elements. 
% Within each main population (P1, Im, P2), [O/Fe] and [Mg/Fe] increase with time, which could reflect enrichment process from core-collapse supernovae (CCSNe).
We also show the relation between [(C+N+O)/Fe] and age in Fig.~\ref{fig:xfe_age_subpops_CNO}. Overall, the trend with age is weak, and apparent increase toward younger ages is largely driven by the most metal-rich population (13) from \ddpayne{}.
This behavior likely reflects the opposite age trends of O relative to C and N, which reduces the net variation when the three elements are combined.
% where we find a less clear enrichment of C+N+O with age older than 11.5~Gyr, after which C+N+O increases rapidly with younger ages.
[Na/Fe] from \ddpayne{} generally shows an enrichment trend with younger age, though the P2 subpopulation 6 appears significantly younger than other groups at the same [Na/Fe] level.
% Another $\alpha$-element [Ca/Fe] shows a rapid increase 
Another $\alpha$-element [Ca/Fe] and the heavy $s$-process element [Ba/Fe] (both from \ddpayne{}) exhibit increases toward younger ages, but with a much larger scatter than the relation with \feh{}. 
Notably, both elements appear to increase rapidly toward younger ages along the sequences defined separately by the Im and P1 subpopulations, with apparent age ranges of $\sim0.13$ Gyr for Im and $\sim0.3$ Gyr for P1.
However, these ranges are comparable to, or smaller than, the intrinsic age spreads reported by \cite{Clontz2026ApJ}, with $\sigma_{\rm age}\sim0.3$--$0.6$~Gyr for the Im and $\sigma_{\rm age}\sim0.24$~Gyr for the P1 subpopulations. Therefore, the trends in the subpopulation medians should be treated with caution.
For [Ca/Fe], the apparent increase could reflect rapid enrichment due to Type II SNe \citep[e.g.,][]{Johnson2010ApJ}.
For [Ba/Fe], the apparent $\sim0.3$~Gyr age range in P1 subpopulations is shorter than the typical lifetimes of low-mass AGB stars of $0.3$--$2$~Gyr \citep{Smith2000AJ}.
If all subpopulations formed within the same environment, such rapid [Ba/Fe] enrichment could indicate additional sources, such as massive rotating stars \citep{Shingles2014ApJ, Frischknecht2016MNRAS}, as suggested by \cite{Dondoglio2026A&A}.

% Put the Milky Way trends on these plots. For [Ba/Fe]
% For [Ba/Fe] increasing with age so fast, could be because the gas reservoir is smaller than the MW, where the ISM is more easily removed due to tidal stripping and stellar winds (SN). Therefore, the AGB mass loss become more effective and make the [Ba/Fe] increase really fast. % \url{https://iopscience.iop.org/article/10.3847/1538-4357/aa978f/pdf}

\subsection{Relative Enrichment of $s$-process and $\alpha$-elements}
\label{subsec:discuss-salpha}

\begin{figure*}[!ht]
    \centering
    \includegraphics[width=1\linewidth]{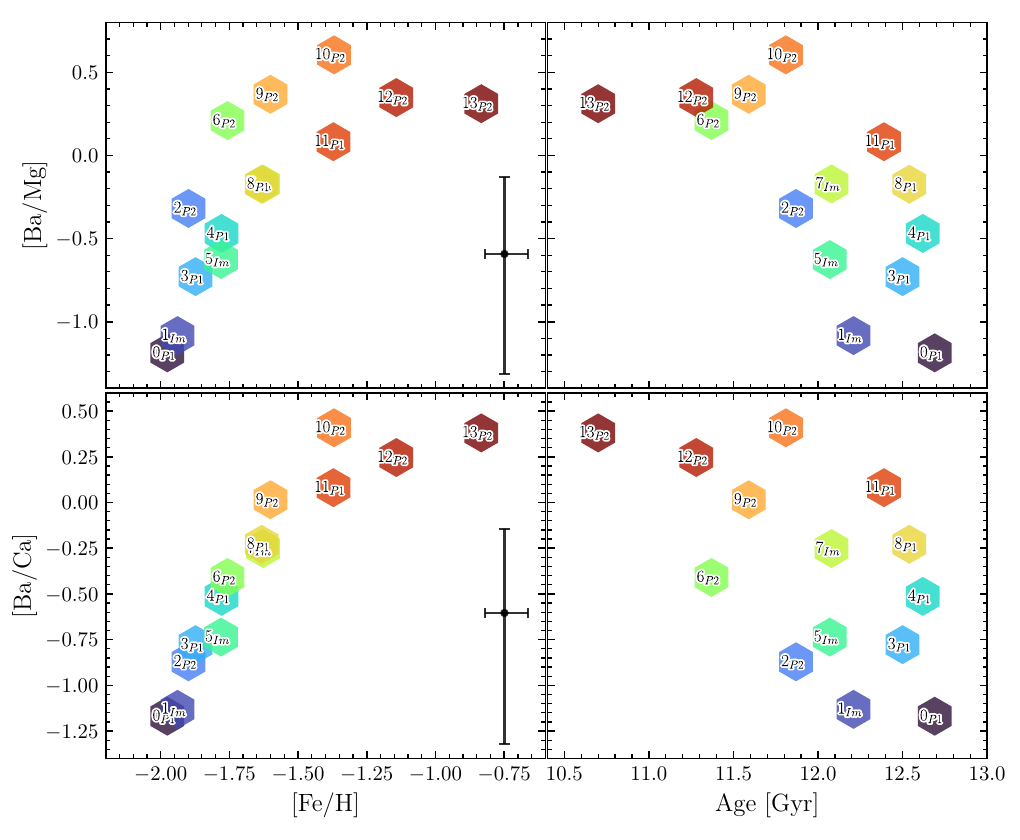}
    \caption{
    Ratios of the heavy $s$-process element Ba to $\alpha$-elements Mg and Ca as a function of \feh{} (left) and age (right).
    Top: [Ba/Mg]; bottom: [Ba/Ca].
    Data points represent median abundances of the 14 subpopulations calculated using [Ba/Fe], [Ca/Fe], and [Mg/Fe] in Table~\ref{tab:subpop_abundances}; symbols and colors are consistent with Fig.~\ref{fig:xfe_xfe_subpops}.
    The black reference error bars show the typical spread in the abundance ratios, calculated by propagating the 16th--84th percentile abundance ranges of the corresponding elements in quadrature for each subpopulation and then taking the median across the subpopulations.
    }
    \label{fig:ba_alpha_feh_age}
\end{figure*}

To further investigate the enrichment history of $\omega$ Cen, we examine the abundance ratios between the heavy $s$-process element Ba and the $\alpha$-elements Mg and Ca. 
In particular, we consider [Ba/Mg] and [Ba/Ca] as functions of both [Fe/H] and age. 
Such ratios have been shown to be sensitive tracers of relative enrichment timescales between neutron-capture and core-collapse supernovae (Type II SNe) products \citep[e.g.,][]{Skuladottir2019A&A}.

The results are shown in Fig.~\ref{fig:ba_alpha_feh_age}. 
We note that [Ba/Ca] is based fully on \ddpayne{} measurements, whereas [Ba/Mg] is partly hybrid because [Mg/Fe] uses literature values for some low-metallicity and Mg-depleted subpopulations.
Compared to [Ba/Fe], which contains Fe in both axes when plotted against [Fe/H], the [Ba/Ca] ratio provides a cleaner diagnostic of relative enrichment.
Because Ca is primarily produced by Type II SNe and does not participate in light-element anti-correlations. 
The [Ba/Ca]-[Fe/H] relation (bottom left of Fig.~\ref{fig:ba_alpha_feh_age}) shows a smooth and continuous increasing trend, reinforcing the picture of progressive $s$-process enrichment with increasing metallicity. 
% This continuity is similar to that seen in [Ba/Fe] but avoids potential covariance introduced by Fe.
However, the [Ba/Mg]-[Fe/H] relation (top left of Fig.~\ref{fig:ba_alpha_feh_age}) shows more complexity due to the Mg-Al anti-correlation. Because Mg is depleted in P2 stars, the [Ba/Mg] ratios of P2 subpopulations appear more enhanced relative to P1 and Im. 
This separation clearly illustrates the impact of light-element variations on chemical enrichment diagnostics and emphasizes the importance of using an $\alpha$-element such as Ca, which is not affected by proton-capture process, when assessing $s$-process enrichment.

The age relations in the right column of Fig.\ref{fig:ba_alpha_feh_age} provide further insight. Similar to the [Ba/Fe] trend in Fig.\ref{fig:xfe_age_subpops_NaMgBa}, both [Ba/Ca] and [Ba/Mg] show an overall increase toward younger ages across the P1, Im, and P2 subpopulations, suggesting relatively more significant $s$-process enrichment compared to $\alpha$-element production. 
The enrichment in the younger subpopulations appears slower than in the older populations. 
However, unlike the tight relations with [Fe/H], the age trends show larger scatter and do not follow a single sequence.
When connecting subpopulations at similar metallicities (e.g., 0-1-2, 3-5-6, 4-7-9), the [Ba/Ca]-age diagram suggests possible evolutionary links consistent with a scenario in which multiple star-forming units at different metallicities experienced internal enrichment before merging into \omc{} \citep[e.g.][]{Clontz2026ApJ}. 
Alternatively, grouping populations by P1, Im, and P2 reveals that older P1 and Im populations exhibit more rapid $s$-process enrichment relative to the younger P2 populations. 
In Section~\ref{sec:discuss-senario}, we combine these abundance relations with \feh{} and age to evaluate the formation scenarios proposed in the literature, to investigate which scenario better matches our results.

\section{Evaluating the Complex Formation Scenarios of Omega Centauri}
\label{sec:discuss-senario}

The combination of our \ddpayne{} abundance measurements with literature data, together with the precise ages and subpopulation classifications from the oMEGACat survey, enables us to evaluate the formation scenarios proposed for \omc{}.
Our main conclusion is that no single existing scenario provides a fully satisfactory explanation of the combined age, metallicity, abundance, and population constraints.
Any scenario must simultaneously explain the observed age distributions, chemical patterns, kinematics, and varied spatial distributions.
In Sections~\ref{subsec:discuss-senario-merge}-\ref{subsec:discuss-senario-multiplesequence}, we discuss three main formation scenarios, highlighting the direct observational evidence from our results and the literature (e.g., chemical abundances, ages, and kinematic constraints) that supports them and that challenges them.
In Section~\ref{subsec:discuss-senario-synthesis}, we summarize the key observational constraints that challenge existing scenarios and then discuss more speculative possibilities, such as IMBH-related processes or tidally triggered events.

\subsection{Scenario 1: Two Environments, Separate Enrichment}
\label{subsec:discuss-senario-merge}

One proposed model is that the P1 and P2 populations originated in separate environments and later merged.
In this scenario, P1 and P2 are not evolutionarily related either because they formed in different locations or because they formed at different times.
This scenario can naturally explain the $\sim$1~Gyr age gap between P1 and P2 \citep{Clontz2024ApJ}, significantly larger than the intrinsic age spreads.
In a single isolated system, it is difficult to maintain a relatively quiescent period with a duration of $\sim1$~Gyr, during which only Im stars were born. However, it can be explained if these two populations are formed at different times in different environments.
\cite{Mason2026MNRAS} found that the Im populations exhibit light-element patterns (such as the Mg-Al anticorrelation) typical of ``normal'' globular clusters. 
Therefore, the Im population could represent stars from an infalling globular cluster.
Spatially, \cite{Sollima2007ApJ}, \cite{Bellini2009A&A}, and \cite{Dondoglio2026A&A} showed that P2/Helium-enriched subpopulations at all metallicities are more centrally concentrated than P1 \citep[but also see][for a more complex picture]{Calamida2017AJ, Calamida2020ApJ}.
This would be an expectation of a merger event where the P2 cluster is a denser cluster and thus ends up more centrally concentrated.

However, our chemical abundance results present challenges to this ``separate origins'' scenario. 
If P1 and P2 formed in completely unrelated systems, their chemical enrichment histories should show no correlation. 
Instead, we observe a remarkably continuous set of chemical trends between the two groups. 
As shown in Fig.~\ref{fig:xfe_feh_subpops_CNO}, Fig.~\ref{fig:xfe_feh_subpops_NaMgBa}, and Fig.~\ref{fig:ba_alpha_feh_age}, [(C+N+O)/Fe], [Ca/Fe], [Ba/Fe] (although some offset is seen between P1 and P2 at high metallicities), and [Ba/Ca] increase smoothly with metallicity rather than forming distinct enrichment tracks. 
We also find a tight correlation between \nfe{} and age (Fig.~\ref{fig:xfe_age_subpops_CNO}). 
It would be very coincidental for the P2 population to continue the $s$-process and C+N+O enrichment from the P1 population. The observed chemical continuity instead suggests that the gas reservoir forming P2 should have been connected to, and chemically influenced by, the evolved stars of the P1 population.

\citet{Mason2026MNRAS} proposed a modified scenario in which P1 formed first in the \omc{} progenitor. The Im population is attributed to inspiralling gas-rich globular clusters, and the gas brought in by these systems subsequently fueled the formation of P2. 
This scenario can still explain the age differences among the P1, Im, and P2 populations, as the introduction of fresh gas and/or stars from external sources (e.g., via mergers) could have restarted star formation and led to the formation of P2 stars. 
It also accounts for the light-element patterns observed in Im stars (e.g., the Mg-Al anticorrelation). 
In this framework, the P1 and P2 populations can remain chemically connected, as suggested by our results, because they formed at the same location; P2 stars would have formed from material polluted by evolved P1 stars and the infalling gas. 
To further verify this scenario, detailed chemical evolution modelling will be required to test whether it can reproduce the abundance trends observed in this study.

\subsection{Scenario 2: Single-Environment Self-Enrichment}
\label{subsec:discuss-senario-selfenrich}

The second scenario suggests that P1, Im, and P2 formed within a single spatially coherent environment, as proposed by \citet{Dondoglio2026A&A}.
In this scenario, \omc{} experienced multiple star-formation episodes, with SNe ejecta being retained to explain the metallicity spread in both P1 and P2.
The P2 population formed directly from gas polluted by the AGB ejecta of massive P1 stars.

If \omc{} was the nucleus of a dwarf galaxy, the deep gravitational potential allows efficient retention of SNe and AGB ejecta.  
This scenario is supported by the chemical continuity with \feh{} and age already discussed in Section~\ref{sec:disscuss}. 
The steady increase and extreme enrichment in $s$-process elements (e.g., [Ba/Fe]$\sim$1.0 dex), as shown in our Fig.~\ref{fig:xfe_feh} and previous studies \citep{Johnson2010ApJ, Marino2011ApJ, Marino2012ApJ}, is consistent with enrichment from low- and intermediate-mass AGB stars (1.5-3~$M_{\odot}$) over timescales of several hundred million years.
There are also tight correlations of [(C+N+O)/Fe], [Ca/Fe], [Ba/Ca] with metallicity and a tight \nfe{}-age relation (Fig.~\ref{fig:xfe_feh_subpops_CNO}, Fig.~\ref{fig:xfe_feh_subpops_NaMgBa}, and Fig.~\ref{fig:ba_alpha_feh_age}).  
Also favoring this scenario are the presence of Na-O and Mg-Al anti-correlations at a range of metallicities (Fig.~\ref{fig:xfe_xfe_subpops}); these relations are similar to those found in typical globular clusters \citep{Gratton2012A&ARv}. 
\citet{Clontz2025ApJ} also found that helium increases with metallicity in the early (metal-poor) P2 sequence, which may be consistent with continuous helium production from massive AGB stars of the P1 population. 
\citet{Dondoglio2026A&A} also cite the concentrated P2 morphology as evidence for this scenario, because centrally retained and diluted ejecta would naturally lead to P2 stars forming preferentially in the cluster core. Nonetheless, this morphology alone cannot rule out a contribution from externally supplied gas (e.g., via inspiralling globular clusters, \citealt{Mason2026MNRAS}).
% although we think the consistent difference at all metallicities in the morphology is more simply explained through the merger of two clusters with different densities, rather than the repeated formation of a P2 population from inflow P1 gas in just the same way.  
%In terms of ages, although the median ages of P1 and P2 differ by $\sim$1 Gyr, the individual stellar distribution along the AMR \citep{Clontz2024ApJ} does not show discrete sequences; instead, it suggests to a continuous, extended star-formation history.

Despite the above supporting evidence, this scenario faces several challenges.  First, it is challenging to explain the large $\sim$1~Gyr separation between the main P1 and P2 star-formation episodes \citep{Clontz2024ApJ, Clontz2026ApJ}.  
Second (and related), it struggles to explain the low metallicities observed in the P2 population (Fig.~\ref{fig:ChM_xfe}). If metallicity enrichment is continuous and P2 formed directly from the processed ejecta of P1, one would expect P2 to begin at a noticeably higher \feh{} than P1. Instead, we observe a significant overlap in \feh{} between P1 and P2, with the most metal-poor stars in P2 only 0.1 dex more metal-rich than the metal-poor tail in P1  \citep{Clontz2026ApJ, Mason2026MNRAS}. Reconciling this metallicity overlap would require substantial dilution of enriched material with pristine gas. This is difficult to achieve without removing key abundance signatures such as the continuous $s$-process and [(C+N+O)/Fe] enrichment and the Na-O and Mg-Al anti-correlations (Fig.~\ref{fig:xfe_feh_subpops_CNO}, Fig.~\ref{fig:xfe_feh_subpops_NaMgBa}, and Fig.~\ref{fig:xfe_xfe_subpops}).
Moreover, a large helium enhancement ($\Delta Y \approx 0.15$) is already present at the lowest metallicity in P2 \citep{Clontz2025ApJ}, which is a challenge to explain with AGB enrichment. %It remains unclear whether AGB stars alone can produce enough helium at such early times to account for this enrichment.
Furthermore, it is unclear why a second star-formation episode (P2) could restart with comparable intensity to P1 after such a long delay, and why the Im population exhibits a noticeably narrower metallicity range \citep{Mason2026MNRAS, Clontz2026ApJ}. \citet{Dondoglio2026A&A} suggested the Im population formed later from a mixture of ejecta from both P1 and P2 stars. However, this scenario would require Im stars to be present at all metallicities. In addition, the SGB ages \citep{Clontz2024ApJ} indicate that the Im population has ages intermediate between P1 and P2, which is inconsistent with this formation pathway.
In terms of subpopulations, Fig.~\ref{fig:xfe_age_subpops_CNO} and Fig.~\ref{fig:xfe_age_subpops_NaMgBa} show that not all the abundance-age relations are monotonic. If P1 and P2 evolved strictly within the same environment, it is hard to explain the complex, non-monotonic age abundance relations seen in these figures, particularly in  [O/Fe], [Na/Fe], and [Mg/Fe]. 
%One possible explanation is the systematics of age determination, but it requires further exploration.
Finally, the most metal-rich populations in both P1 and P2 (populations 11 and 13 in Fig.~\ref{fig:xfe_feh_subpops_CNO}, Fig.~\ref{fig:xfe_feh_subpops_NaMgBa}, Fig.~\ref{fig:xfe_age_subpops_CNO}, and Fig.~\ref{fig:xfe_age_subpops_NaMgBa}) appear as outliers relative to the overall system. Their distinct chemistry is difficult to reconcile with a single-environment, purely self-enriched formation pathway.

\subsection{Scenario 3: Multiple Environments, Concurrent Enrichment}
\label{subsec:discuss-senario-multiplesequence}

A third possibility, suggested by the ``two-stream'' age-metallicity relation identified by \citet{Clontz2024ApJ}, is that \omc{} represents a complex assembly of multiple star-forming units (e.g., subpopulations 0-1-2, 3-5-6, 4-7-9 in Fig.~\ref{fig:xfe_feh_subpops_CNO}) that evolved in parallel before merging. 
In this scenario, each major \feh{} group underwent its own localized chemical enrichment within the central regions of a dwarf galaxy but in slightly different localized environments (e.g., a central nucleus vs. surrounding clumps).
This scenario attempts to bridge the gap between the first two. 
It naturally accommodates the tight age-metallicity sequence seen in Fig. 4 of \citet{Clontz2024ApJ} and also explains the large metallicity spread at a given age and the overlap of metal-poor tail between P1 and P2.
Moreover, it alleviates the requirement for strictly monotonic enrichment in the abundance-age relations (Fig.~\ref{fig:xfe_age_subpops_CNO} and Fig.~\ref{fig:xfe_age_subpops_NaMgBa}).
The small differences in light-elemental abundances and anti-correlations at different metallicities could be due to each group experiencing its own globular-cluster-like enrichment cycle independently and creating its own ``second-generation'' stars.
% \textcolor{red}{After these star-forming units merge, star formation can continue, leading to further chemical enrichment (e.g., subpopulations 8, 11, 12, and 13).}

However, like Scenario 1, this scenario raises questions regarding the apparent chemical continuity between the different metallicity groups (Fig.~\ref{fig:xfe_feh_subpops_CNO}, Fig.~\ref{fig:xfe_feh_subpops_NaMgBa}).
% this scenario also faces the ``chemical coincidence'' problem. 
Although combined abundance trends with [Fe/H] from all the star-forming units can be relatively tight if formed in the same dwarf galaxy \citep[e.g.,][]{Shetrone2026ApJ}, 
the continuous and extended chromosome sequences of P1 and P2 \citep{Clontz2025ApJ} suggest a level of coordinated enrichment that may be difficult to reconcile with independent evolution of multiple star-forming units.
Instead, they can be more easily explained if the metal-rich populations (whether in P1 or P2) formed from the core-collapse supernovae ejecta of the metal-poor populations within the same sequence.
Moreover, 
the extreme helium enhancement observed across all P2 populations \citep{Clontz2025ApJ} is not typically seen in independent and less massive systems.
% It is difficult to explain why different metallicity groups exhibit continuous abundance variations along the P1 and P2 sequences in the chromosome diagram (as shown in Fig.~\ref{fig:ChM_xfe}), and why P2 is consistently helium-enhanced to a level beyond that in typical globular clusters \citep{Clontz2025ApJ}. 
% These behaviors are not naturally explained if each metallicity group formed in an independent environment. 
Furthermore, this scenario does not account for all identified subpopulations. In particular, the absence of Im stars at the highest metallicities leaves populations 10, 11, 12, and 13 without a clear evolutionary connection.

\subsection{Summary of All the Scenarios}
\label{subsec:discuss-senario-synthesis}

As detailed in the previous sections, no single formation scenario presented here or in the literature provides a full explanation for the complex formation history of \omc{}. 
By combining the observational evidence from our \ddpayne{} and literature results, we identify five key factors that make the interpretation of \omc{}'s formation history challenging:
\begin{enumerate}
    \item The large age gap ($\sim1$~Gyr) between two main star formation episodes P1 and P2.% with the similarly intense star formation.
    \item The significant metallicity overlap between P1 and P2, with Im in between.
    \item The continuous chemical enrichment in N, total C+N+O, Ca, and Ba across the P1, Im, and P2 populations. %and the Na-O and Mg-Al anti-correlations, with the exception of the most metal-rich P1 subpopulation. 
    \item The extreme helium enrichment in the P2 population even at its lowest metallicities. 
    \item The origin of Im populations with relatively narrower metallicity spreads compared to P1 and P2.
\end{enumerate}

One possible interpretation of the age gap is that the later populations (Im and/or P2) were triggered by close passages or disk crossings of the progenitor dwarf galaxy through the Milky Way. 
In this scenario, tidal interactions compress or funnel gas toward the central regions, inducing new separated star-formation episodes. After repeated interactions, the outer regions of the dwarf galaxy would be stripped, leaving only the remnant nucleus observed today as \omc{}. This scenario preserves a common nuclear environment while allowing externally triggered star formation, and such tidally induced nuclear starbursts have been demonstrated in numerical simulations \citep[e.g.,][]{Bekki2003MNRAS}.

In addition, several other important factors have not been considered. 
From a kinematic perspective, \citet{Cordoni2020ApJ} and \citet{Ziliotto2026A&A} showed that P2 stars exhibit higher radial anisotropy than P1 stars. This feature can be interpreted in multiple ways. P2 may have formed from infalling material or an accreted subsystem, or it may reflect gas ejected from the P1 population falling back toward the cluster center with slightly different kinematic properties.
However, any present-day kinematic differences are not necessarily primordial, as radial anisotropy can be imprinted during the subsequent long-term dynamical evolution of the cluster \citep[e.g.,][]{Tiongco2019MNRAS, Aros2025A&A}.
The recent confirmation of an intermediate-mass black hole (IMBH) in the center of \omc{} \citep{Haberle2024Natur} may also introduce additional complexity, although its role in the chemical evolution remains unclear.
The IMBH may provide a missing link to the extreme helium enrichment of P2 \citep{Clontz2025ApJ}, potentially through events such as the formation of supermassive stars (SMS) \citep[e.g.][]{Gieles2025MNRAS} or long-lived ``immortal'' stars in an AGN-like disk that happened between the formation of P1 and P2 \citep[e.g.][]{Dittmann2023ApJ}, or the formation and rejuvenation of massive stars via dynamical interactions in the dense cluster core \citep[e.g.,][]{Rantala2025MNRAS}. 
The IMBH could further influence chemical enrichment indirectly by modifying the retention, mixing, and redistribution of stellar ejecta in the central regions.

In the future, combining the chemical abundances presented here with oMEGACat proper motions and MUSE radial velocities \citepalias{Haberle2024ApJ} will enable chemodynamical studies that may provide further insight into the formation and assembly history of \omc{}.

% The IMBH could further influence the chemical enrichment of other elements.
% Moreover, measurements of C and N abundances remain challenging, and systematic uncertainties in these elements could affect the inferred continuity of C+N+O (M. Mackenzie, private communication).

% the extreme helium enrichment ($\Delta Y \approx 0.15$) observed in P2 stars (Clontz et al. 2025b) requires a pollution source capable of processing hydrogen at high temperatures, such as supermassive stars (SMS) or fast-rotating massive stars (FRMS). The recent confirmation of an intermediate-mass black hole in $\omega$ Cen (Häberle et al. 2024b) may provide a link to the SMS channel, as these stars are invoked as potential seeds for IMBHs.

\section{Summary} 
\label{sec:summary}

In this work, we apply the neural network model \ddpayne{} to MUSE spectra of 7,302 RGB stars in the oMEGACat dataset.
We derived chemical abundances of [C/Fe], [N/Fe], [O/Fe] at the metal-rich end ($\feh{}_\mathrm{uncal}>-1.0$~dex), [Mg/Fe] for $\feh{}_\mathrm{uncal}>-1.5$~dex, and [Na/Fe], [Ca/Fe], and [Ba/Fe] across the full metallicity range of \omc{}.
By combining our measurements with high-resolution spectroscopic literature, we constructed the most comprehensive chemical abundance distributions of \omc{} on the chromosome diagram to date (Fig.~\ref{fig:ChM_xfe}).
We investigated abundance trends with metallicity and age by separating stars into the widely adopted P1, Intermediate, and P2 populations (Fig.~\ref{fig:xfe_feh_stream}), as well as the 14 subpopulations identified by \cite{Clontz2026ApJ} (Fig.~\ref{fig:xfe_feh_subpops_CNO}-\ref{fig:ba_alpha_feh_age}).
Our main results are summarized as follows:
\begin{enumerate} 
    \item P2 stars are consistently enhanced in [N/Fe] and [Na/Fe] and depleted in [O/Fe] and [Mg/Fe] compared to P1 stars at similar metallicities.
    \item The Im populations occupy an intermediate position between P1 and P2 in chemical space but largely follow the chemical trends of P1.
    \item The total [(C+N+O)/Fe], $\alpha$-element [Ca/Fe], and the $s$-process element [Ba/Fe] increase continuously with metallicity across the P1 and P2 populations, suggesting a related formation environment. 
    \item P2 stars show a clear [Ba/Fe] offset relative to the P1 and Intermediate populations.
    \item Analysis of the 14 subpopulations reveals a strong correlation between [N/Fe] and stellar age, suggesting nitrogen acts as a tracer for the cluster's enrichment history. [Ba/Fe] also exhibits a weaker dependence on age, consistent with delayed enrichment from low- and intermediate-mass AGB stars.
    \item The [Ba/Ca] ratio increases smoothly with metallicity and shows an overall increase toward younger ages, indicating relatively stronger neutron-capture enrichment relative to core-collapse supernovae production.
\end{enumerate}

Furthermore, we evaluated three proposed formation scenarios for \omc{} using observational evidence of chemical abundances, spatial distributions, kinematics, and ages. 
This analysis demonstrates the difficulties in reconciling their constraints to identify the definitive formation history of \omc{}.

Our study demonstrates that machine learning approaches applied to integral-field spectroscopy can effectively recover detailed chemical abundances in crowded environments where traditional multi-object spectroscopy is limited. 
The resulting dataset provides valuable new constraints for chemical evolution models aiming to reconstruct the formation history of \omc{}.

% \section{Future Studies} 
% \label{sec:future}

% Apply this to other globular clusters and 
The application of \ddpayne{} to \omc{} in this work demonstrates the feasibility of measuring chemical abundances from MUSE spectra in crowded stellar fields. 
This method can be extended to the wider MUSE globular cluster survey \citep{Kamann2018MNRAS}, which includes observations of the central regions of 25 clusters. 
Analyzing these datasets will provide spatially resolved abundance maps and provide essential constraints on the chemical patterns of multiple populations.

% Better understanding on stellar yields to resolve the formation history of omega cen
% To resolve the formation puzzle of \omc{} and better interpret the chemical abundance patterns and their relation to age, more precise stellar nucleosynthesis yields are required. 
A detailed comparison between the observed abundance trends and theoretical stellar yields will be necessary to fully constrain the enrichment pathways of helium, light, and $s$-process elements.
In particular, updated yields for massive stars \citep[e.g.,][]{Limongi2018ApJS} and AGB stars at low metallicities with different masses \citep[e.g.,][]{Karakas2014PASA} and their enrichment timescales are needed to better constrain the enrichment for the helium, light, and $s$-process elements as discussed in Section \ref{sec:discuss-senario}.

% Better DDPayne
Moreover, the abundance measurements can be improved with updates to the \ddpayne{} model. 
New training sets using APOGEE DR19 \citep{SDSSCollaboration2026ApJS} and GALAH DR4 \citep{Buder2025PASA} will provide more precise abundance measurements and better coverage of metal-poor stars ($[\mathrm{Fe/H}] < -1.5$). 
The upgraded \textsc{TransformerPayne} \citep{Rozanski2025ApJ} model could also provide more robust and accurate abundance estimates from MUSE spectra.

%% Also note that the akcnowlodgment environment does not support long amounts of text. If you have a lot of people and institutions to acknowledge, do not use this command. Instead, create a new \section{Acknowledgments}.
\begin{acknowledgments}
We thank C. Foster, M. R. Hayden, C. Manea, S. Martell, M. McKenzie, and J. van de Sande for useful discussions.
Z.W., A.C.S., and C.C. acknowledge the support from a \textit{Hubble Space Telescope} grant GO-16777. AFK acknowledges funding from the Austrian Science Fund (FWF) [grant DOI 10.55776/ESP542].
M.A.C. acknowledges the support from FONDECYT Postdoctorado project No. 3230727.
This work is done using \texttt{Yoga} (\url{https://yoga-server.github.io/}), a privately built Linux server for astronomical computing.
\end{acknowledgments}


\vspace{5mm}
\facilities{HST(STScI), VLT:Yepun (MUSE)}


\software{Astropy \citep{AstropyCollaboration2013A&A, AstropyCollaboration2018AJ, AstropyCollaboration2022ApJ}, Numpy \citep{Harris2020Natur}, Scipy \citep{Virtanen2020NaMet}, Matplotlib \citep{Hunter2007CSE}, \ddpayne{} \citep{Ting2017ApJL, Xiang2019ApJS}}



% \appendix

% \section{Chromosome Diagram with Literature Chemical Abundances}
% \label{appsec:ChM-abundance-lit}

% \begin{figure*}[!ht]
%     \centering
%     \includegraphics[width=1\linewidth]{ChM_lit_xfe_snr50_mask_mpoor.png}
%     \caption{
%     Same as Fig.~\ref{fig:ChM_xfe}, but overplotted with individual stellar abundance measurements from literature studies, as indicated in the top-right corner of each panel.
%     }
%     \label{fig:ChM_xfe_lit}
% \end{figure*}
% Fe/H: good training set range coverage
% Mg/Fe: due to systematics with teff from APOGEE-Payne


\bibliography{paper}{}
\bibliographystyle{aasjournalv7}



\end{CJK*}
\end{document}